\documentclass[12pt]{iopart}
\usepackage{graphicx}
\usepackage{rotating}
\usepackage[usenames]{color}

\def\onlinecite{\cite}

\begin{document}

\title{Novel behaviors of monolayer quantum gases on Graphene, Graphane and Fluorographene}

\author{Luciano Reatto$^1$, Davide E. Galli$^1$,
Marco Nava$^{1}$~\footnote{Present address: Computational Science,
Department of Chemistry and Applied Biosciences, ETH Zurich,
USI Campus, Via Giuseppe Buffi 13, CH-6900 Lugano, Switzerland},
and Milton W. Cole$^2$}

\address{
$^1$ Dipartimento di Fisica, Universit\'a degli Studi di Milano, via Celoria 16, 20133 Milano, Italy \\
$^2$ Department of Physics, Penn State University, University Park, PA 16802 USA
}
\ead{luciano.reatto@mi.infn.it}
\begin{abstract}
This article discusses the behavior of submonolayer quantum films (He and H$_2$) on graphene 
and newly discovered surfaces that are derived from graphene. 
Among these substrates are graphane (abbreviated GH), which has an H atom bonded to each C atom, 
and fluorographene (GF). 
The subject is introduced by describing the related problem of monolayer films on graphite. 
For that case, extensive experimental and theoretical investigations have 
revealed that the phase diagrams of the bose gases $^4$He and para-H$_2$ are qualitatively similar, 
differing primarily in a higher characteristic temperature 
scale for H$_2$ than for He. The phase behaviors of these films on one side of pristine graphene, 
or both sides of free-standing graphene, 
are expected to be similar to those on graphite.
We point out the possibility of novel phenomena in adsorption on graphene related to the large 
flexibility of the graphene sheet,
to the non--negligible interaction between atoms adsorbed on opposite sides of the sheet and to 
the perturbation effect of the 
adsorbed layer on the Dirac electrons.
In contrast, the behaviors predicted on GF and GH surfaces are very different from those on graphite, 
a result of the different corrugation, i.e., the lateral variation of the potential experienced by 
these gases. This arises because on GF, for example, 
half of the F atoms are located above the C plane while the other half are below this plane. 
Hence, the He and H$_2$ gases experience a very different potentials 
from those on graphite or graphene. As a result of this novel geometry and potential, distinct 
properties are observed. 
For example, the $^4$He film's ground-state on graphite is a two-dimensional (2D) crystal 
commensurate with the substrate, 
the famous $\sqrt{3}\times\sqrt{3}$ R30$^o$ phase; on GF and GH, instead, it is predicted to 
be an anisotropic superfluid. 
On GF the anisotropy is so extreme that the roton excitations are very anisotropic, as if the 
bosons are moving in a multiconnected space along the bonds of a honeycomb lattice. 
Such a novel phase has not been predicted or observed previously on any substrate. 
Also, in the case of $^3$He the film's ground-state is a fluid, thus offering the 
possibility of studying an anisotropic Fermi fluid with a tunable density. 
The exotic properties expected for these films are discussed along with proposed experimental tests.

\end{abstract}

\maketitle

\section{Introduction}
The properties of monolayer films are core subjects within the fundamental sciences of chemistry 
and physics, as well as the more applied fields of chemical engineering, 
electrical engineering and materials science. From a basic science perspective, this subject is 
enriched by the wide range of phenomena exhibited by the various substrate/film 
combinations \cite{ref1}. This article reviews a subset of these problems - quantum films on 
graphene-related substrates at low temperature. For such systems, 
quite novel behaviors have been predicted, with many open questions to address. 
This article describes both problems that are relatively well understood 
(i.e., films on graphite) and those that have just begun to be investigated. 
We restrict the focus to helium and hydrogen films because these exhibit similar, 
intriguing behaviors in spite of having rather different bulk phase properties. 
We note at the outset that uncertainty about interaction models exists even for the problem of
adsorption on graphite; the newly studied systems -graphene and its derivatives- present
significantly greater uncertainty in the interactions.

The subject of monolayer quantum films has an intriguing history. Considerable attention has 
been devoted to a highly idealized problem: the helium film in 
mathematical two-dimensions (2D), an appropriate model if the third dimension has no significant role. 
This model was a logical subject of investigation 
because superfluidity (SF) and Bose-Einstein condensation (BEC) are fundamentally important 
subjects that have been extensively investigated in 3D. 
From that research, there has emerged a consensus, beginning with the work of Fritz London, 
that BEC and SF are causally related properties of bose fluids in 3D. 
Given this background, particular interest in the subject of He monolayer films was piqued 
by an apparent contradiction of this consensus view: 
$^4$He films down to a few layer thickness were known experimentally to exhibit SF, while 
rigorous theorems demonstrated that BEC cannot exist 
in such films at any nonzero temperature \cite{ref2} $T$. The resolution of this paradox first 
emerged in the 1970's, with the introduction, 
by Kosterlitz and Thouless and by Berezinski, of the concept of topological 
long range order (TLRO) \cite{ref3,ref4}. 
This view affirms that while infinite long range order (of the superfluid order parameter) 
associated with BEC does not exist in $^4$He films at finite $T$, 
the decay with distance (an algebraic function) of the two-point correlation function of the 
local order parameter is sufficiently slow that such very long 
range correlations suffice to yield SF. 

This TLRO hypothesis is more than qualitative; theory predicts that the superfluid 
density has a jump from a finite value $\rho_s$ to zero at 
the transition temperature $T_c$ and yields an explicit relationship for the ratio $\rho_s/T_c$ , 
with $T_c$ being determined by an instability 
of the superfluid with respect to unbinding of quantized vortices \cite{ref4}, a 
prediction that has been confirmed experimentally in multilayer 
$^4$He films \cite{ref5}. The nature of this 2D transition is thus completely different from that 
of the 3D SF transition (for which $\rho_s$ vanishes in a continuous way at 
$T_c$ with a critical behavior in the universality class of the XY model). In 3D, the transition 
temperature is conventionally described by the phenomenological 
Landau model, in which the value of $T_c$ is derived from the spectrum of elementary excitations, 
and by a complementary view, which identifies the transition 
as the point when BEC disappears. 

One may question the relevance of the idealized 2D model of a monolayer film, insofar as the 
model ignores the substrate, apart from its nominal role 
of confining the atoms to a plane. Indeed, the experimental behavior of most monolayer films is 
significantly affected by the substrate surface's structure. 
In addition, superfluidity disappears altogether when the film thickness becomes one or two layers, 
depending on the substrate- the so-called ``inert layer''. 
Many experimental studies have been carried out on disordered materials, in which case the 
resulting behavior can be quite complicated and different 
from that expected to occur on ordered surfaces. In fact, there are relatively few studies 
to date of quantum monolayers on well-characterized, flat surfaces. 
The principal reason for this is experimental: thermodynamic, structural and scattering probes 
of films require "high surface area" materials, 
meaning those with a high ratio $(A/V)_{sub}$ of exposed surface area $A$ to volume $V$. 
This constraint is particularly acute for the quantum films, 
He and H$_2$, since many conventional surface techniques (e.g., STM, LEED, AFM) are not feasible 
with these very weakly bound films because the probe 
itself alters these films. Fortunately, there is at least one class of materials (layered materials) 
which provide both high surface area and ordered, 
planar surface facets. Graphite is the most studied member of this group. For example, one 
commercially available form of porous carbon, 
Grafoil, possesses specific area of order 20 m$^2$ per gram and presents exposed graphitic 
facets of characteristic lateral extent greater than 100 \AA. 
This fortunate situation is a consequence of the anisotropy of graphite, i.e. the strong 
intra-planar binding compared to the weak inter-planar bonding. 
This anisotropy lets extended facets survive the treatment used to prepare the substrate in 
high surface area form.
Since the 2D density ($\rho$) of monolayer He or H$_2$ on this surface is $\rho \simeq 0.12$ atoms/\AA$^2$, 
one gram of substrate (corresponding to $N_C = 5 \times 10^{22}$ atoms of carbon and surface area 
$A=20$ m$^2$) yields a relatively high ratio of monolayer He atoms or H$_2$ molecules to substrate C atoms: 
$N_{{\rm He}({\rm H}_2)}/N_{\rm C} = \rho A/N_{\rm C} \simeq 5 \times 10^{-3}$. 
The relative contribution of the film to the thermal properties is significantly enhanced 
above this ratio because the thermal ``background'' contribution from the 
graphite substrate is relatively small at low $T$. Thus, the total specific heat can 
possess a measurably large contribution from a submonolayer 
film even though the film represents a minute fraction of the atoms present.

\begin{figure}[h]
\begin{center}
\includegraphics*[width=12cm]{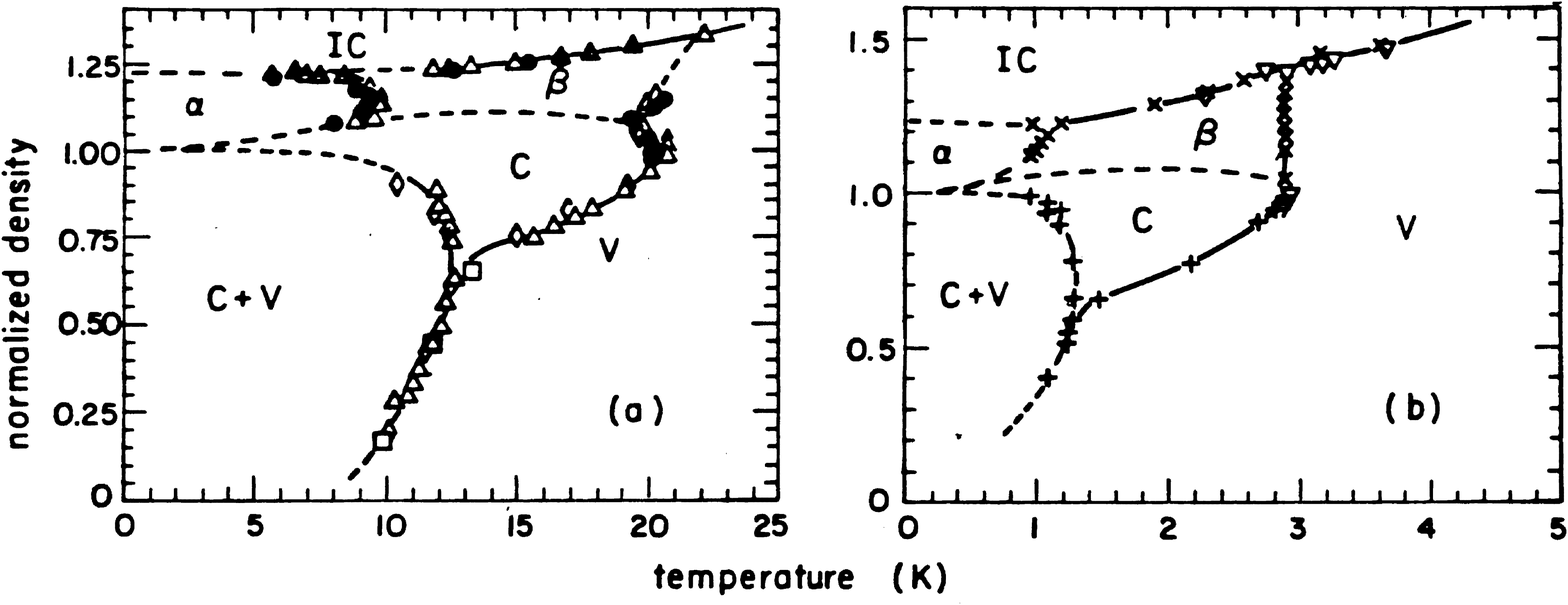}
\caption{
Phase diagrams of (a) parahydrogen and (b) $^4$He on graphite, from Motteler \cite{ref9}, 
adapted by Bruch, Cole and Zaremba \cite{ref6}. 
The ``normalized density'' is 2D film density $\rho$ relative to $\rho_C$. 
These diagrams are derived from heat capacity experiments of many
groups \cite{ref6,ref7,ref8,ref9,ref10,ref11,ref12}. C (IC) denotes commensurate (incommensurate) 
solid and V denotes vapor; $\alpha$ and $\beta$ are conjectured 
to be striped incommensurate and reentrant fluid phases, respectively.
}\label{c1f1}
\end{center}
\end{figure}

The importance of the substrate in understanding quantum films became evident with the detailed 
exploration of the phases of He and H$_2$ on graphite, 
beginning in the late 1960's. This work has been the subject of several reviews \cite{ref6,ref7,ref8}. 
Phase diagrams for these systems \cite{ref9} are shown in figure \ref{c1f1}. 
The ``normalized density'' label of the ordinate scale refers to the ratio of the actual 2D 
density $\rho$ to that of a commensurate phase ($\rho_C \sim 0.0637$ atoms/\AA$^2$). 
The latter phase, denoted C, is seen to be a prominent feature of both He and H$_2$ monolayers. 
It is a $\sqrt{3} \times \sqrt{3}$ R30$^o$ phase, for which the adatoms 
occupy 1/3 of the strong binding sites, localized above the centers of the graphite hexagons. 
This C phase is also present for $^3$He with an order-disorder transition 
at a temperature $T=3$ K, essentially identical to that for $^4$He.

Given such intriguing properties of the quantum films on graphite, one may wonder about the properties 
to be found for such films on related materials. 
The present paper explores this problem, with a focus on graphene and two graphene-derived materials, 
fluorographene (abbreviated GF) \cite{ref14} and graphane (GH) \cite{ref13,ref15}.
These new substrates are obtained from graphene by chemically bonding F or H, respectively, to the C 
atoms.

The title of this article includes the words ``novel behaviors''.
Let us summarize what these words refer to, even though space
considerations limit our discussion of some of these topics:
\begin{enumerate}
\item The strongest binding adsorption potential for these gases;
\item The most corrugated potential, with a novel periodicity;
\item An unprecedented anisotropic monolayer liquid ground state,
with an anisotropic roton spectrum and a normal fluid constituting 40\% of the total mass at zero temperature
(The possibility of a non--zero normal component at $T=0$ K was envisioned, 
in principle, but has not yet been observed. See: \cite{new0a,new0b});
\item An unusual wetting transition on graphene, due to the interaction
between adsorbed molecules on opposite sides of free-standing graphene \cite{new1};
\item Anticipated large mutual elastic interaction between adatoms, due to the flexibility of graphene \cite{newB};
\item Unique controllable ``tunability'', with adsorption, of the novel electronic states of graphite \cite{new2};
\item Possible novel existence of a two-dimensional 3He liquid ground state;
\item Imbibition transition, in which gas lifts a graphene sheet off of a silica substrate \cite{new3};
\item The possibility of two alternative commensurate structures of these films on free-standing 
graphene.
\end{enumerate}

The next section reviews our current understanding of monolayer adsorption on graphite. 
In Section \ref{sec3}, we discuss the differences between that behavior and that 
expected on a free-standing graphene substrate. Sections \ref{sec4} and \ref{sec5} describe the very 
unusual properties that have been predicted recently for the 
quantum films on GF and GH. To the best of our knowledge, no experiments have tested these predictions 
thus far. We hope that the exotic behaviors that we describe here 
will stimulate such investigations. Our conclusions and outlook for the future are contained in 
Section \ref{sec6}.

\section{Monolayer quantum films on Graphite}\label{sec2}

Figure \ref{c1f1} exhibits an intriguing variety of phases for $^4$He and H$_2$. Such behavior is 
completely different from what would occur for the idealized 2D version 
of these phase diagrams, which would be qualitatively similar to that of 3D He, differing by the 
absence of commensurate phases and those denoted $\alpha$ and 
$\beta$ in the figure. There remains some controversy about aspects of the monolayer behavior 
(i.e., the phase boundaries and their interpretations), 
a situation we shall not address here \cite{ref16,ref17,ref18,ref18b,ref19,ref20,ref21}. Instead, we will 
describe and interpret the principal features of the 
various monolayer phases, for which there does exist a consensus opinion. 

At low $\rho$ and $T>1$ K, the He film behaves like a weakly interacting 2D gas, with small, but
observable, quantum--statistical effects. This behavior was first seen in experiments of the 
groups of Dash and Vilches at the University of Washington (UW) and Goodstein at Caltech \cite{ref9,ref16}.
In particular, the high T specific heat reached the classical 2D limit ($C/N=k_B$). 
Siddon and Schick demonstrated for both $^3$He and $^4$He isotopes that the experimental specific 
heat data above $T=1.5$ K can be understood in terms of a 
2D quantum virial expansion \cite{ref22}, in which the  virial coefficients 
(distinct values of $B_2(T)$ for the two isotopes due to different spin 
and therefore statistics) were calculated from a Lennard-Jones (LJ) interatomic potential, 

\begin{eqnarray}\label{lj}
V_{LJ}(r) = 4\epsilon \left[ \left(\frac{\sigma}{r}\right)^{12} - 
\left(\frac{\sigma}{r}\right)^{6} \right]	
\end{eqnarray}
Their 2D analysis employed the same interaction parameters ($\epsilon/k_B=10.2 K$, $\sigma=$2.56 \AA) 
which are consistent (approximately) 
with the thermodynamic and structural behavior of 3D helium. The agreement between that virial 
expansion and the experimental data of the UW group \cite{ref9} 
indicated that the low density behavior was quantitatively described by such a 2D model. 
This finding was very encouraging, insofar as no adsorption 
system had ever demonstrated such consistency with a 2D theory prior to this work. 
Refinements of the Siddon-Schick analysis were proposed subsequently, 
including (a) the use of substrate screening of the pair potential, 
a $\sim$ 10\% effect \cite{ref23,ref24}, and (b) the demonstration that the 
single particle properties exhibit band structure effects arising from the 
periodic adsorption potential \cite{ref25,ref26}.

\begin{figure}[h]
\begin{center}
\begin{minipage}{6cm}
\includegraphics*[width=6cm]{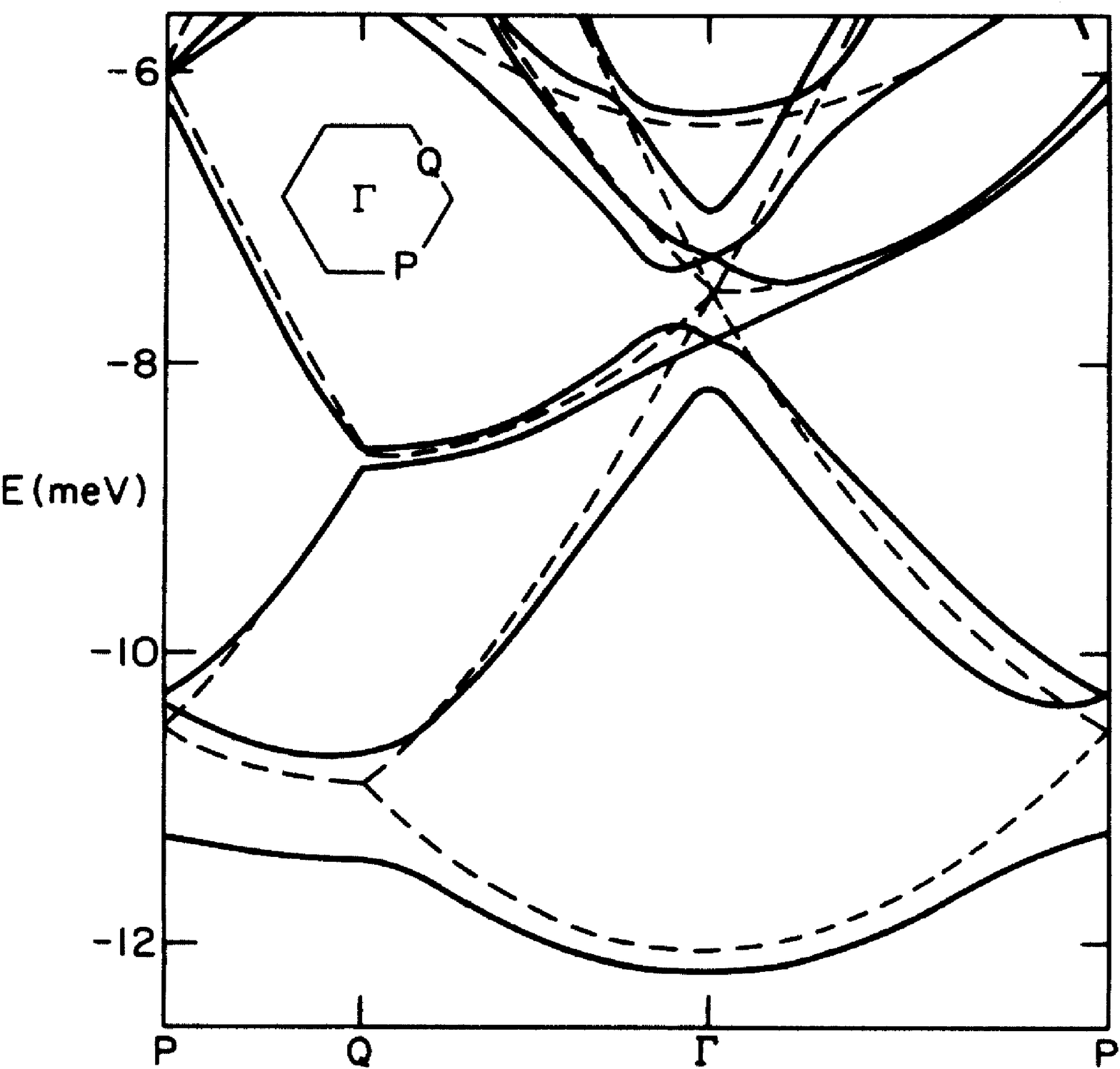}
\end{minipage}\hspace{1cm}
\begin{minipage}{6cm}
\includegraphics*[width=6cm]{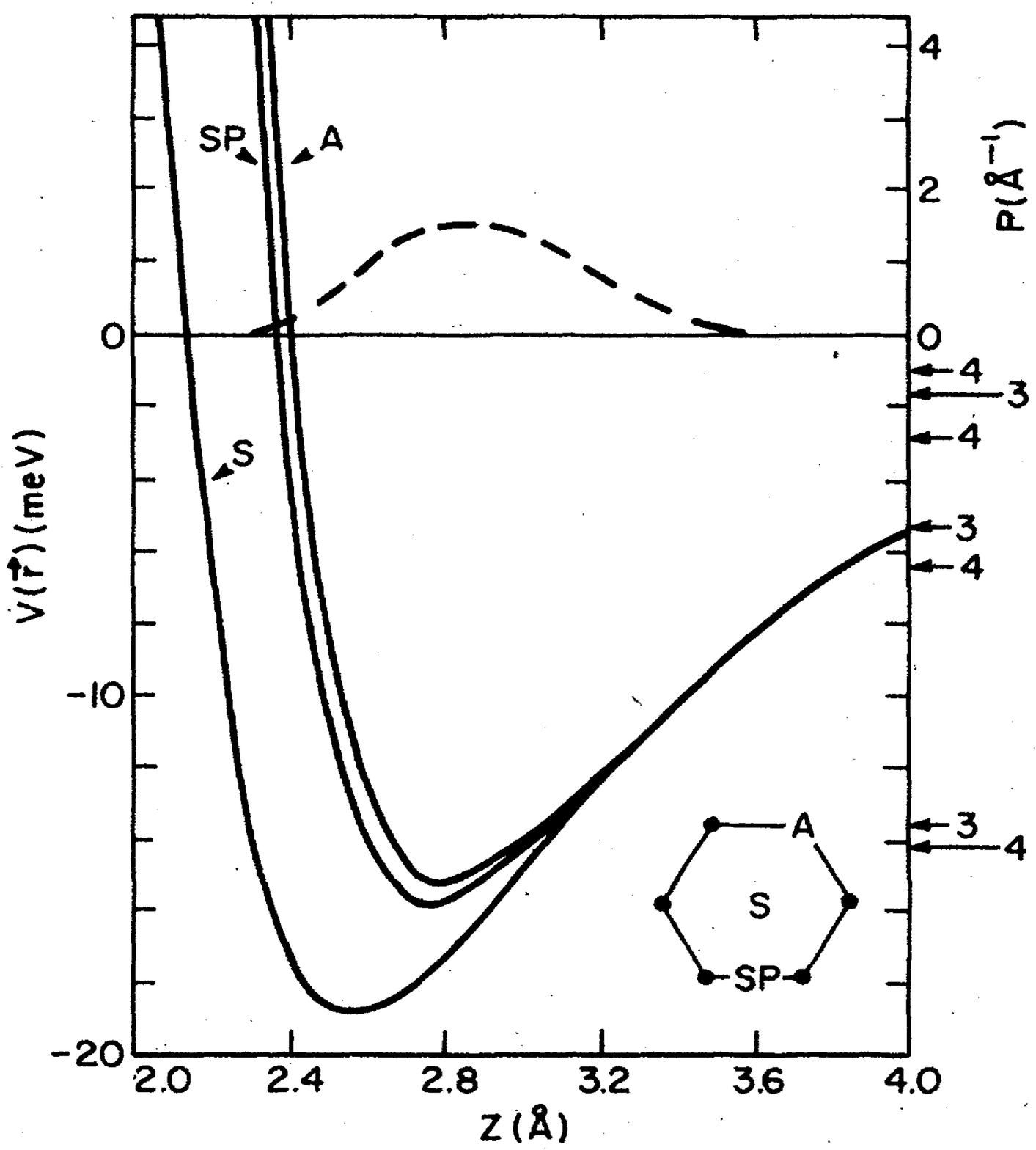}
\end{minipage}\caption{\label{c1f2}
Left: Band structure of $^4$He on graphite (full curve) for wave vector $\vec{k}$ along symmetry 
lines in the 2D Brillouin zone (shown). 
Dashed curve is the free-particle result, omitting any corrugation terms \cite{ref25}. 
Right: Model potential $V(x,y,z)$ and ground-state wave function (dashed curve) for a $^4$He atom
on graphite \cite{ref25}, derived from vibrational energy levels (right scale)
measured with He/graphite 
scattering \cite{ref28,ref29} S, A and SP labels denote indicated $(x,y)$ positions for 
the potential curves. 1 meV $\simeq$ 11.6 K.
}
\end{center}
\end{figure}

Fig. \ref{c1f2} shows results of semiempirical calculations of the adsorption potential 
of $^4$He/graphite and the 
corresponding band structure \cite{ref25,ref26,ref27}.
One observes that the lateral variation of the potential well-depth across the surface is of 
order 15\% and the band gaps induced by this corrugation are of order 
0.2 meV $\sim$ 2 K. The effective mass enhancement for $^4$He is modest:
$m^{\star}/m_4 = 1.06$ ($m_4$ is the bare mass). 
These findings might be described as ``weak corrugation'', since the lowest band width 
is much larger ($\sim$ 10 K) than the gap. That fact explains why the strictly 2D virial 
analysis is essentially consistent with the experimental data at low $\rho$. 
However, there does exist one dramatic consequence of the corrugation, the commensurate phase, 
which is seen to dominate the monolayer phase diagrams of both gases. 
It may seem paradoxical that the corrugation potential can be neglected in one regime of coverage 
but have a significant effect in another regime. 
One way to rationalize these quite distinct behaviors is to recognize that the thermal de Broglie 
wavelength for these particles $\lambda=\left[2\pi\hbar^2/mk_BT\right]^{1/2}$ 
is large compared to the spacing ($\sim$ 2.5 \AA) between hexagons at low temperature $T$; 
a He atom (mass $\sim$ 4 amu), has $\lambda \sim $ 9 \AA~ at $T=1$ K,
four times the lattice constant. While single atoms therefore manifest weak effects of the 
corrugation, many-particle systems show large effects- the C phases. 
The analysis of this latter problem is necessarily subtle, requiring detailed quantum statistical 
calculations. This subtlety is evident from the observed ``reentrant'' 
role of the corrugation potential, as a function of atomic diameter $\sigma$. 
While the corrugation is critically important for the quantum films
(characterized by a small value of $\sigma$),
the effect is essentially negligible for Ar, with an intermediate size $\sigma$,
but the corrugation becomes important again for gases (CH$_4$, Kr and Xe) with larger values of $\sigma$.

\begin{figure}[h]
\begin{center}
\includegraphics*[width=12cm]{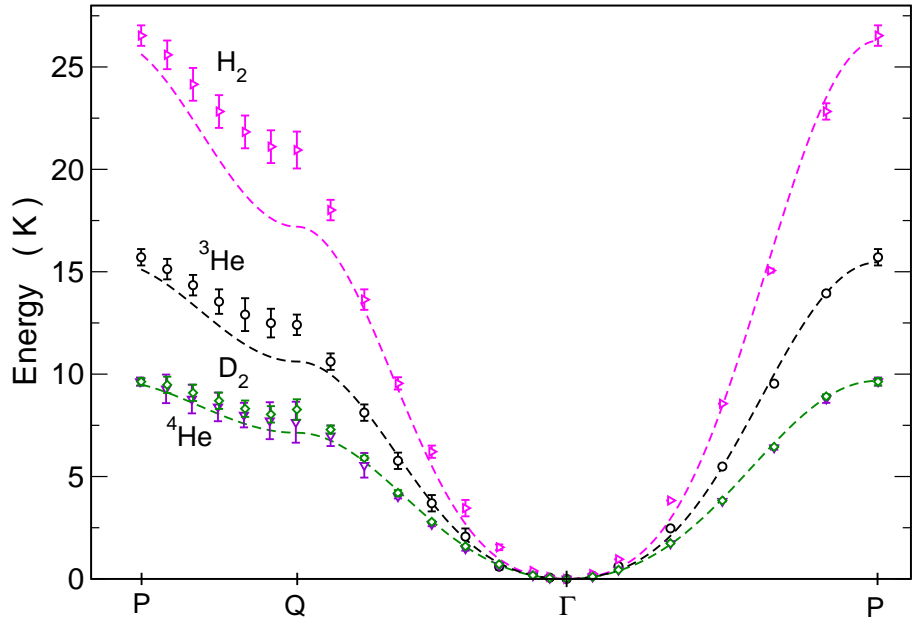}
\caption{(Color online)
Energy bands computed for various gases on graphite (symbols), and the fits to the tight-binding model 
discussed in the text (dashes).
}\label{c1f3}
\end{center}
\end{figure}

The band structure in Fig. \ref{c1f2} is based on energy band theory and that computation takes 
into account a large but finite number of Fourier components of the adsorption potential. 
We have computed the lowest excited state of the single $^4$He, $^3$He and H$_2$ as function of 
the $x-y$ wave vector $\vec{k}$ by a quantum Monte Carlo method 
described in the next section. As adsorption potential for He we have used the same potential 
as in ref.~\onlinecite{ref25} and for H$_2$ the potential from 
ref.~\onlinecite{ref50}. The lowest energy bands along some of the principal directions are 
shown in Fig.~\ref{c1f3} together with a fit 
based on a nearest neighbor (n.n.) and a next-nearest-neighbor (n.n.n.) tight binding model:     
\begin{eqnarray}                                 
E(k_x,k_y) = 6(t_1-t_2)+ \\
+2t_2\left[\cos(3lk_x)+\cos(1.5lk_x+1.5\sqrt{3}lk_y)+\cos(1.5lk_x-1.5\sqrt{3}lk_y)\right]- \nonumber \\
-2t_1\left[\cos(1.5lk_x+0.5\sqrt{3}lk_y)+\cos(1.5lk_x-0.5\sqrt{3}lk_y)+\cos(\sqrt{3}lk_y)\right] \nonumber
\end{eqnarray}
Here $t_1$ and $t_2$ are the nearest neighbor and next-nearest neighbor coupling constants 
and $l=1.42$ \AA~ is the distance between two carbon atoms.
The parameters $t_1$ and $t_2$ have been obtained
by fitting the band structure dispersion data displayed in Fig.~\ref{c1f3}
and take the following values (in Kelvin units): 
for $^3$He $t_1$ = 1.70(1), $t_2$ = 0.37(1). For $^4$He and D$_2$: $t_1$=1.08(1), 
$t_2$ = 0.18(1) and for H$_2$ $t_1$=2.77(2), $t_2$ = 0.63(2).

The band energy of H$_2$ and of D$_2$ on graphite have not been evaluated previously. 
In table \ref{c1tab1} the effective mass $m^{\star}$ at the $\Gamma$ point are given and in 
the case of the He isotopes compared with those given by the energy band theory \cite{ref25}.
Note that none of these mass enhancements is large, consistent with the wide 
bands and the delocalized single particle wave functions.

\begin{figure}[h]
\begin{center}
\includegraphics*[width=12cm]{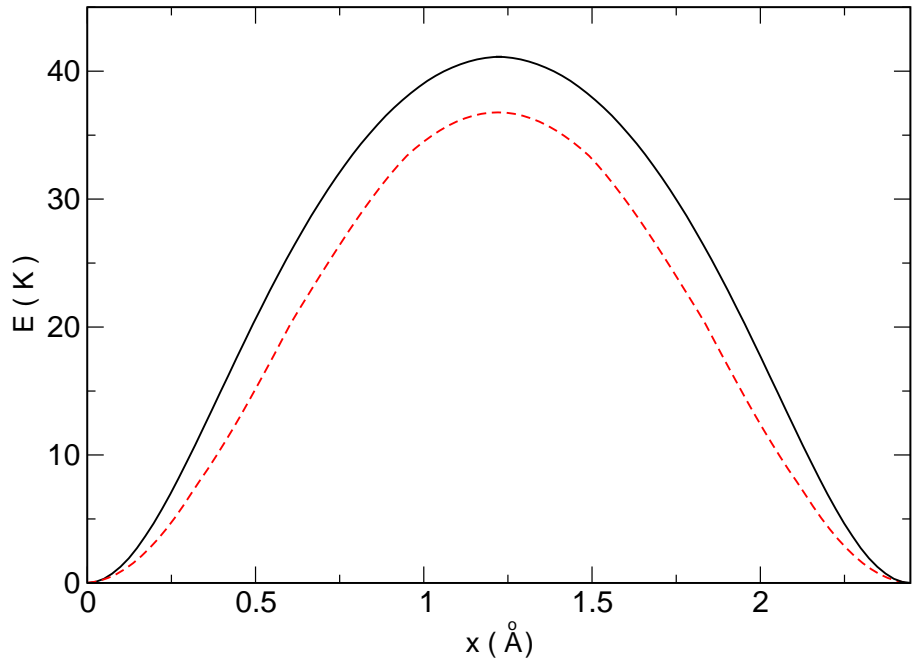}
\caption{(Color online)
Potential energy of He (full line) and H$_2$ (dashed line) as a function of position along a 
line connecting two hexagon centers, 
separated by 2.46 \AA. The zero of energy is shifted to facilitate comparison. 
This energy represents the minimum, 
as a function of $z$, for each lateral position $(x,y)$ across the surface.
}\label{c1f4}
\end{center}
\end{figure}

Note in Fig.~\ref{c1f3} the near coincidence of the band structures of $^4$He and D$_2$. 
This behavior is a result of the near compensation of two differences: 
the greater mass $^4$He of is accompanied by a somewhat greater curvature 
(and potential barrier to translation), as seen in Fig. \ref{c1f4}, so the effective frequency 
of low energy oscillation is similar to that of D$_2$.

\begin{table}[h]
\caption{\label{c1tab1} Band mass enhancement $m^{\star}/m$ for various gases on 
graphite (unpublished); values in parentheses were computed previously from empirical scattering
data \cite{ref25}}

\begin{center}
\begin{tabular}{ | c | c |}
\hline
  Adsorbate &$m^{\star}/m$\\
\hline
H$_2$ & 1.03 \\
D$_2$ & 1.10\\
$^3$He & 1.08 (1.04) \\
$^4$He & 1.10 (1.06) \\

\hline

\end{tabular}
\end{center}
\end{table}

\begin{figure}[h]
\begin{center}
\includegraphics*[width=12cm]{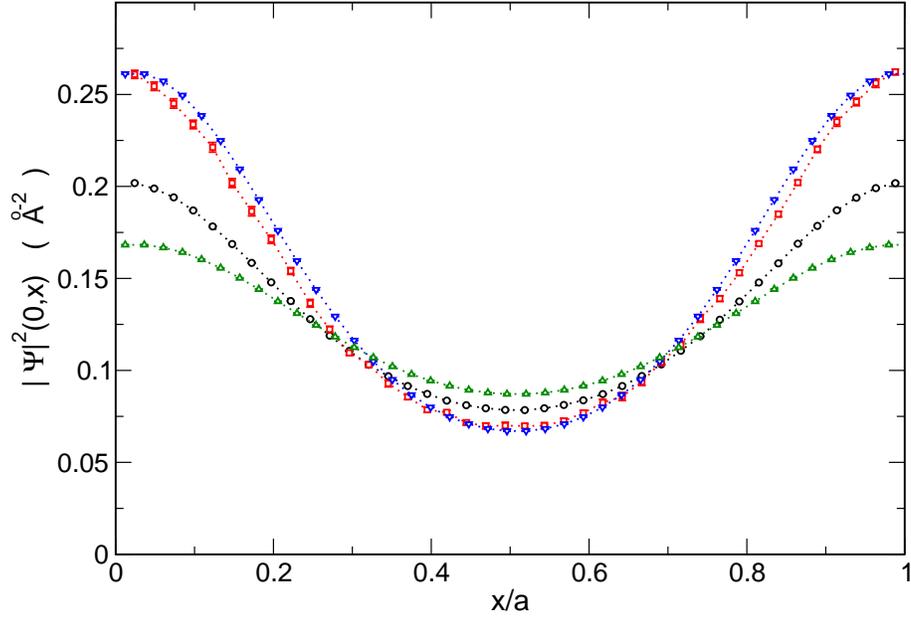}
\caption{
(Color online) Square of the ground state wave function along the direction connecting 
neighboring adsorption sites for a single atom on graphite. The plotted lines 
correspond to $^4$He (red Squares), $^3$He (black Circles), H$_2$ (green Triangles up), 
D$_2$ (blue Triangles down). 
}\label{c1f5}
\end{center}
\end{figure}

In Fig. \ref{c1f5} the square of the ground state wave function is plotted along the direction 
connecting neighboring adsorption sites for $^3$He, $^4$He, H$_2$ and D$_2$.
Again one can notice the similarity of $|\Psi|^2$ of $^4$He and D$_2$.

At intermediate coverage, the phase diagrams of these quantum particles in Fig. \ref{c1f1} 
are seen to be dominated by the $\sqrt{3}\times\sqrt{3}$ R30$^o$ phase. 
One of the very interesting features of the associated order-disorder transition 
(melting of the commensurate solid) involves the critical temperature $T_c$ 
of this transition \cite{ref30,ref31}. First, we note that the value  of $T_c$ is virtually 
independent of isotopic species. 
That is, $T_c \sim 3$ K for both $^3$He and $^4$He while $T_c \sim 20$ K is common to 
both H$_2$ and D$_2$. This isotope-independence is striking since 
the zero-point energy (ZPE) differs by a factor $1/3$ in the two He cases
and the ZPE usually plays a prominent role in the physics of condensed quantum fluids;
the factor is even larger (1/2) for H$_2$.
The explanation of the mass-independence of $T_c$ in the present case is the dominant role 
played by the highly corrugated adsorption potential. 
That contribution to the energy localizes the He atoms (H$_2$ molecules), to an extent 
apparently indifferent to the mass, and this localization energy is 
overcome only by thermal energy, with ZPE playing a relatively small role. 
The ratio (20 K)/(3 K)$ \sim $ 7 of the two disordering temperatures 
is presumably due to the significantly larger mutual interaction potential in the 
hydrogen case than for He.

\begin{figure}[h]
\begin{center}
\includegraphics*[width=12cm]{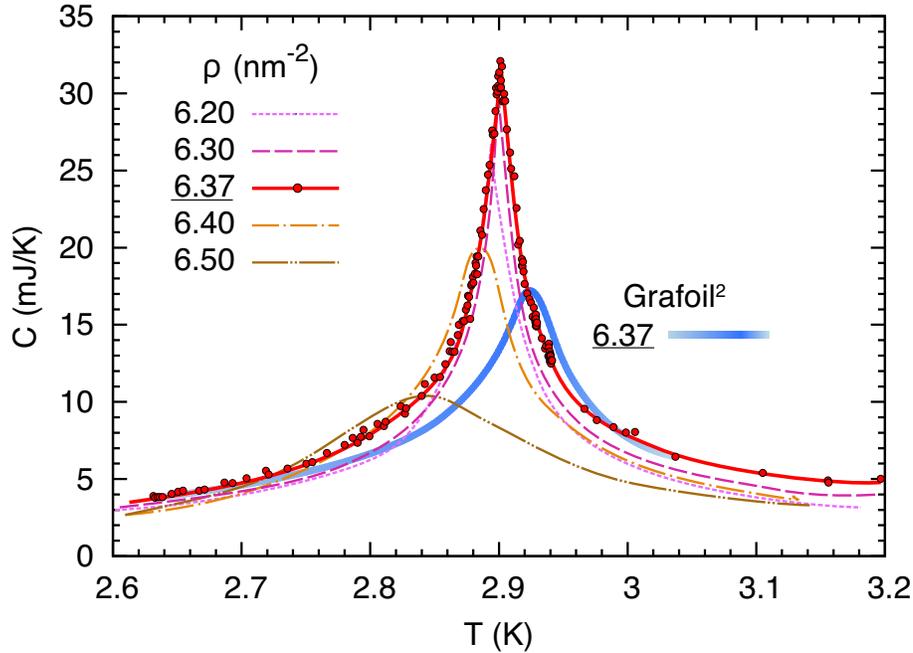}
\caption{(Color online)
Heat capacity of $^4$He on ZYX graphite, at indicated coverages near the critical density
0.0637 \AA$^{-2}$ and temperature $T_c$ ($\sim 3$ K) from Nakamura, et al. \cite{ref32};
there also appears one curve for $^4$He/Grafoil, from Greywall \cite{ref18b};
see also Ecke et al. \cite{ref33}. On this surface (area 30.5 m$^2$), 
a classical ideal gas would have $C=2.68$ mJ/K
}\label{c1f6}
\end{center}
\end{figure}

Most of the phase transition behavior attributed to these adsorbates has been identified by and 
characterized with the specific heat data. 
One significant example is the commensurate solid melting transition, for which the specific 
heat $C(N,T)$ of $^4$He is shown in Fig. \ref{c1f6} 
for two varieties of graphite substrate \cite{ref32,ref18,ref33}. 
The dramatic peaks in the data rise a factor 10 above the value expected for a 
2D classical ideal gas: $C/(Nk_B)=1$, i.e. $C \sim 2 $ cal/(mol-K). 
When first seen in 1971 \cite{ref30}. this transition's dramatic signature was the most 
unequivocal thermodynamic evidence of a phase transition in any submonolayer film. 
Its prominence in the $N-T$ plane helps to confirm the commensurate 
film coverage which is determined by the spacing (4.26 \AA) between second neighbor 
adsorption sites on graphite. 
(Neighboring sites are too close, 2.46 \AA~ apart, to become occupied, given the hard-core 
diameter $\sigma \sim$ 2.6 \AA~ of both He and H$_2$.) 
As seen in Fig. \ref{c1f6}, the divergent behavior of the specific heat is most striking at the 
corresponding density, $\rho_C$=0.0637 \AA$^{-2}$. 
The $N$ and $T$ dependences of $C(N,T)$ were found to be consistent with the 
3-state Potts model, in which the atoms occupy a sublattice 
corresponding to 1/3 of the available sites. That model, proposed by Alexander \cite{ref34}, 
groups adsorption sites into sets of three;
a variable (1, 2 or 3) is then determined by which specific site (among the three possible 
sublattices on graphite) is occupied among the group of three sites. 
The film's total energy within this model is determined by the interaction between 
neighboring groups, which is determined by where the relevant atoms are located. 
The 3-state Potts model is analogous to the 2-state Potts model, i.e., the Ising model, 
for which there are two choices for the variable characterizing 
the site occupation; Ising behavior was observed for $^4$He/Kr/graphite by 
Tejwani et al. \cite{ref35}, for which the geometry corresponds exactly to that of the 
Ising model (with two equivalent sets of adsorption sites). 
Within the spirit of modern phase transition theory, the
Ising and Potts models capture the essential physics of their respective transitions, 
because they incorporate the correct symmetry of each problem. 
Even though the short-range interaction is oversimplified and the long-range interaction is 
completely omitted, the critical behavior of these models arises from long-range correlations, 
which are accurately described by the models. 
Specifically, the 3-state Potts model predicts that near the transition, 
the heat capacity $C(T)$ is proportional to $|T-Tc|^{-\alpha}$, where $T_c$ 
is the critical temperature and $\alpha$=1/3. The agreement with the measured 
critical value ($\alpha=0.34\pm0.01$) for this transition on graphite represents a 
major success for both theory and experiment \cite{ref32,ref33,ref36}.

Fig.~\ref{c1f6} also demonstrates the sensitivity of the thermodynamic measurements to surface quality,
especially near the critical transition. The data are shown for both Grafoil and an alternative
material, ZYX Graphite. The ZYX facets are an order of magnitude larger than those of Grafoil,
but the specific area (2 m$^2$ per gram) of ZYX is an order of magnitude smaller;
the price paid by this much lower specific area of ZYX is not a problem when analyzing the
singularity, since the background specific heat of the substrate is a smooth function of $T$.
One observes in
Fig. \ref{c1f6} that the full-width at half maximum of the sharpest peak on ZYX is about 0.05 K, one-half
the width on Grafoil. The narrower critical region and higher peak value in the ZYX data reflect
the relative absence of inhomogeneous broadening; the correlation length diverges at $T_c$ and one
comes closer to model ideal with ZYX than with Grafoil.

We turn next to the high density phase, labeled IC in figure \ref{c1f1}, seen with the 
approach of the coverage to monolayer 
completion ($\rho_m \sim 0.12$ atoms/\AA$^{2}$ for $^4$He and H$_2$, while $\rho_m \sim 0.11$ 
atoms/\AA$^2$ for $^3$He). 
This full monolayer phase is an incommensurate 2D solid for both helium and hydrogen. We note that the very 
high density reflects a factor 
of 3 compression relative to the 2D ground state phase density ($\rho\sim 0.04$ atoms/\AA$^2$). 
There are two reasons for 
this large compression. One is simply that these 2D quantum fluids are highly compressible. 
For example, the  sound speed of 2D 
$^4$He is predicted \cite{ref36b} to be $s_{2D} \sim 80 m/s$, about 1/3 of the value for 3D $^4$He.

The other reason for the huge compression of the monolayer involves the substrate potential.
Note that the monolayer coverage $\rho_m$ represents the point when particles 
begin to occupy a second layer of the film. 
This is also the point when the coverage increases rapidly as a function of chemical potential.
In the case of adsorption on graphite, $\rho_m$ is a particularly high coverage because the 
potential binding the particles to the 
surface is so attractive; the well-depth $\sim 16$ meV $\sim 180$ K (Fig. \ref{c1f2}) greatly 
exceeds all other energies in the problem. 
In fact, this value was once believed to be more attractive than the adsorption potential 
for He on any other surface. That statement, however, 
now needs to be modified, according to our group's recent results on the GF surface, 
discussed in Section \ref{sec4} below.

An intriguing region of these phase diagrams ($\alpha$ and $\beta$ phases) is confined 
within a narrow coverage domain near $\rho =1.1 \rho_c$. 
The phase(s) in this region are variously interpreted as a striped incommensurate phase 
or a reentrant fluid phase. 
Although its nature is not yet resolved conclusively, the narrow peaks in the heat 
capacity data \cite{ref37} leave no doubt that there is such a 
transition, occurring near 1K for He and 10 K for H$_2$.
 
From the theoretical point of view most of the ab--initio studies
for given atom--substrate and atom--atom potentials deal with Bosons in 
mathematical 2D \cite{dued1,ref41b} or in a translationally invariant adsorption potential \cite{smo1}.
This modeling is not realistic for He or H$_2$ on graphite due to the strong corrugation of the
adsorption potential. For instance the ground state of $^4$He for such models is a superfluid
and not a commensurate non--superfluid state as shown by experiment.
Only when a realistic representation of the corrugated adsorption potential is used one finds
good agremment with experiments.
Only few exact microscopic quantum simulations of submonolayer films of
$^4$He and H$_2$ on graphite taking into account
the effect of the corrugation for the subtrate potentials have been performed.
The simulations were carried out at finite temperature
with the path integral monte carlo (PIMC) method \cite{corr1,corr2,ref19,corr3,ref20}
and at zero temperature with the Diffusion Monte Carlo method \cite{ref53}.
First $^4$He on graphite was studied finding the $\sqrt{3}\times\sqrt{3}$ R30$^o$ commensurate phase 
as the stable phase at low coverages \cite{corr1};
at a coverage below 1/3 these simulations provided evidence that the low-density monolayer 
consists of solid clusters coexisting
with a low--density vapor. The decisive role of the substrate corrugation in stabilizing
the commensurate 1/3 phase was assessed as well as the presence of an incommensurate
triangular solid phase was proved in a subsequent
publication by the same group \cite{corr2}. More recently, new accurate PIMC simulations have enriched
the monolayer phase diagram finding, at higher coverage than the 1/3 phase, a new commensurate (7/16)
phase before the incommensurate solid phase \cite{ref20}; the two commensurate phases were found separated by a
domain--wall phase.
The role of the substrate corrugation for submonolayer molecular hydrogen on graphite
was investigated in ref.~\cite{ref19}. At low temperature the phase diagram was found qualitatively similar to the one
of $^4$He. The transition to an incommensurate solid was studied in a later work \cite{corr3}.

Note that the phase diagrams presented above show no liquid-vapor transition for either 
H$_2$ or $^4$He. Such a condensation transition is 
predicted by a strictly 2D theory \cite{ref38}, but this transition can be preempted by the 
commensurate phase on bare graphite. 
Greywall, however, \cite{ref18} has interpreted his heat capacity data for 
$^4$He on graphite as showing a liquid-vapor transition 
at low coverage, contrary to Fig.~\ref{c1f1}.
This evidence and interpretation are not supported by 
any other experiments. On the theoretical side, 
numerous calculations for $^4$He on graphite agree with the phase diagram of Fig.~\ref{c1f1} 
in the sense that there exists no liquid-vapor transition. 
Instead, at low $T$ and low $\rho$ the commensurate solid is predicted to coexist with a 
low-density vapor \cite{corr1,corr2}. In principle, the presence, 
or absence, of a liquid-vapor transition at coverage below that of the commensurate state 
depends crucially on the corrugation of the adsorption potential. 
Above a threshold corrugation, the C phase is stable relative to the liquid, a conclusion 
consistent with nearly all of the experiments and calculations for graphite.

\begin{figure}[h]
\begin{center}
\includegraphics*[width=12cm]{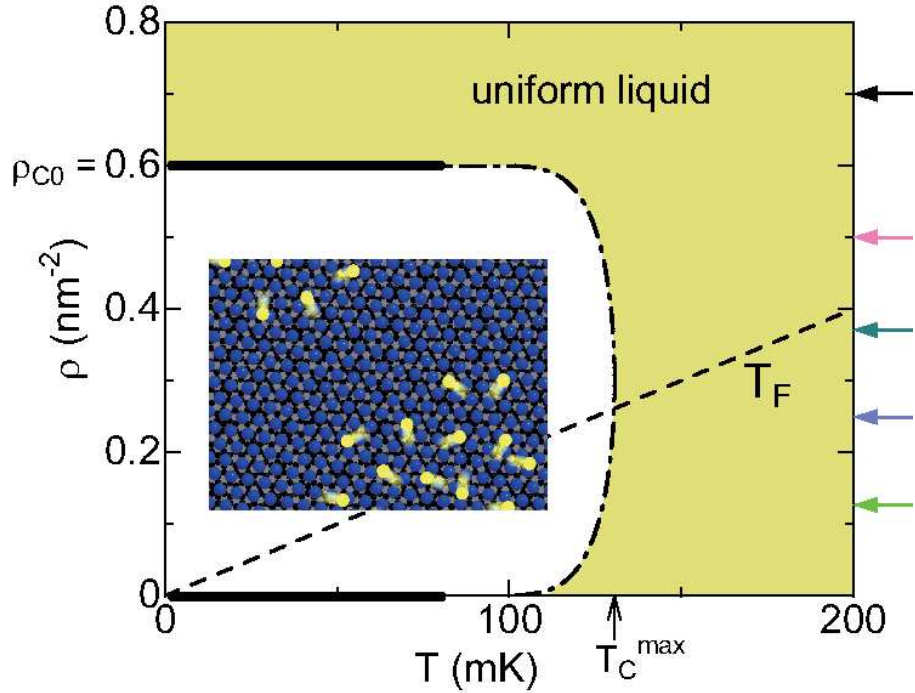}
\caption{(Color online)
Very low density region of the phase diagram of Sato et al. \cite{ref39}, showing the 
condensation transition of $^3$He on graphite. 
Arrows indicate densities at which measurements were made, with solid lines denoting 
transitions determined experimentally, 
while the dash-dot line represents a normalized 2D transition curve. 
Inset depicts 2D puddles schematically within the two-phase coexistence region. 
Dashed line represents the Fermi temperature $T_F$ of an ideal 2D $^3$He gas.
}\label{c1f7}
\end{center}
\end{figure}

The case of $^3$He at low density may provide a contrasting phase behavior to that of $^4$He. 
Note, first of all, that the 2D ground state of $^3$He (unlike $^4$He) 
is predicted to be a gas; that is, there is no condensation transition \cite{ref38b,ref38c}. 
What about $^3$He on graphite? There is evidence for this isotope that 
a very low temperature condensation transition does occur at remarkably low density. 
The transition behavior deduced from specific heat measurements by 
Sato et al \cite{ref39} appears in Fig.~\ref{c1f7}, \cite{ref40} with speculation 
concerning the details of the coexistence curve, especially the critical 
temperature (denoted $T_{cmax}$ in the figure).  The $^3$He liquid ground state is 
seen to have density $\rho_{c_0}\sim 0.006$~\AA$^{-2}$, 
1/6 that of $^4$He; the mean interparticle spacing at the critical density is quite 
large: $\sim (\pi\rho_{c_0}/2)^{-1/2} \sim 10$ \AA,
reflecting the marginal binding of this phase.

These very recent observations would appear to be inconsistent with calculations showing 
that a $^3$He gas does not condense in a 
strictly 2D world \cite{ref38b,ref38c}, a consequence of quantum statistics since a 2D bose fluid of 
(imaginary) He atoms with mass 3 amu would 
be expected to condense \cite{ref38b}. However, on the basis of an approximate 
variational computation it has been argued \cite{ref41} 
that the out-of-plane motion of the film reduces the condensed phase energy, so that the 
ground state on graphite is a self-bound quasi-2D liquid. 
This result has not yet been confirmed by more advanced QMC methods like those used to 
study $^3$He in 2D \cite{ref38b,ref38c}. 
Note in this figure the very low values of the ground state density, 0.006 \AA$^{-2}$. 
Comparison can be made with the corresponding quantity 
for $^3$He in 3D: $\rho_{c_0}^{3D}\sim ~0.0083$~\AA$^{-3}$; scaling the latter quantity 
in a plausible way yields a nominal ``scaled'' 2D density 
$(\rho_{c_0}^{3D})^{2/3}=0.041$~\AA$^{-2}$. The order of magnitude lower values on the 
graphite surface reflect the very weak binding of this quasi-2D 
$^3$He fluid (compared to the 3D case). Another basis for comparison is the behavior 
predicted \cite{ref41b} 
for the critical point of $^4$He in 2D: $\rho_{c_0} \sim 0.025$~\AA$^{-2}$ and $T_c \sim 0.9$ K, manifestly 
higher values than for $^3$He. 
With respect to the presence of a liquid-vapor coexistence at very low coverage we notice 
that the size of the graphitic facets could have a 
significant role.
With an interparticle distance of order 10 \AA~ on a facet of lateral extent of order 100 \AA~
typical of Grafoil, on each facet the number of He atoms is of order 100 and no sharp
transition can be expected with such small number of atoms.
Should a liquid--vapour transition exist for $^4$He on graphite as for $^3$He this should be
seen only on graphite material with much larger facets like ZYX graphite.

\section{Quantum films on Graphene}\label{sec3}
\subsection{General comments}
Most treatments of He or H$_2$ adsorption on graphite evaluate the potential energy, 
of an adparticle at position $\vec{r}$, as a pairwise sum of 
contributions from the C atoms, at positions $\lbrace R_j\rbrace$:
\begin{eqnarray}\label{eqgrp}
V_{graphite}(\vec{r}) = \sum_j U(\vec{r}-\vec{R}_j)
\end{eqnarray}

Here $U(\vec{x})$ is the two-body interaction between a C atom and the adparticle. 
H$_2$ is often treated as though it is one spherically symmetric particle,
although such an approach omits the anisotropy of the molecule. 
This is appropriate at low $T$ because the H$_2$ molecules are in the $L=0$ 
rotational ground state and the adsorption potential does not include any large 
electrostatic terms, which would couple to the molecular orientation.~\footnote{
in the case of H$_2$ in a nanotube, there has been predicted to occur a significant
breaking of the rotational symmetry, in agreement with neutron scattering data which
reported a breaking of the L-manifold degeneracy.
See D. G. Narehood, M. K. Kostov, P. C. Eklund, M. W. Cole, and P. E. Sokol,
``Deep Inelastic Neutron Scattering of H$_2$ in Single--Walled Carbon Nanotubes'',
{\em Phys. Rev. B} {\bf 65}, 233401 (2002).
} 
Eq. (\ref{eqgrp}) can be rewritten as a sum over graphitic layers, with the top 
layer providing the adsorption potential ($V_{graphene}$) on graphene. 
Within this 2-body approximation, Eq. (\ref{eqgrp}), the difference 
$\Delta V=V_{graphene} - V_{graphite}$ between adsorption potentials on graphene 
and graphite corresponds to the net contribution from all layers below the exposed top facet. 
For the case of the quantum gases, the equilibrium distance 
above graphite is $\langle z \rangle \sim 3 $~\AA \cite{ref45}, which means that $\Delta V$ 
corresponds to the interaction energy contributed by a 
graphite half-space located beyond a perpendicular distance $z_{\perp} \cong (3+3.3)$~\AA~= 6.3 \AA~ 
from the adatom. Because $z_{\perp}$ is so large, 
this difference potential $\Delta V$ involves only attractive interactions and is nearly 
invariant with respect to $x/y$ translation of the adparticle. 
Thus, the lateral variation (``corrugation'') experienced by a monolayer film is essentially 
identical for the graphene ($V_{graphene}$) and graphite ($V_{graphite}$) cases. 
Representing the ``missing'' half-space as a continuum, at distance $z_{\perp}$ from the adatom, 
we can approximate the monolayer potential on 
graphene with the relation derived from summing the attractive part of Eq. (\ref{eqgrp}) 
over the missing substrate layers, 
each of which produces an interaction proportional to the inverse fourth power of distance:
\begin{eqnarray}\label{eqgrp2}
\Delta V = V_{graphene}(\vec{r})-V_{graphite}(\vec{r}) \cong 3 C_3 \delta
\sum_{n=0}\frac{1}{(z_{\perp}+n\delta)^4}\cong
\frac{C_3}{(z_{\perp}-\frac{\delta}{2})^3} \quad .
\end{eqnarray}
The 1/2 in the right-most expression corresponds to the fact that asymptotically,
the potential corresponds to a continuum solid, bounded by a plane lying one--half
lattice constant outward from the top layer of ions.
Here $C_3$ is the coefficient of the atom (molecule)-graphite van der Waals interaction, 
$\delta=3.34$~\AA, the interlayer spacing of graphite, 
and we have used the Euler-MacLaurin summation formula. Theoretical values \cite{ref44} are 
$C_3=180 (520)$ meV-\AA$^3$ for He (H$_2$). 
Since $z_{\perp}$ is large compared to the atom's rms fluctuation normal to the surface, 
$\sqrt{\langle (\Delta z_{\perp})^2 \rangle} \sim 0.25$~\AA, 
Eq. (\ref{eqgrp2}) represents a nearly constant shift of the monolayer energy per 
particle on graphene, compared to that on graphite. 
From the second term in Eq. (\ref{eqgrp2}), these shifts are 21 K (62 K) for He (H$_2$), 
which are of order one-tenth of the well-depths of these 
adsorption potentials on graphite \cite{ref46}. A consequence of this constant shift is 
that the energy bands of the adsorbed atom or molecule on graphene are 
negligibly different from those on graphite.

Before discussing the results to date concerning adsorption on graphene, we mention other 
differences between this substrate and graphite. 
One involves the importance of the phonon-mediated interaction between two He atoms (H$_2$ molecules). 
This problem has been explored for atoms 
on graphite and found to be small \cite{ref47}. One cannot be certain, however, that the same 
result would be found on graphene. 
In particular, the graphene case involves a much more flexible substrate than graphite, 
suggesting that the so-called ``mattress effect'' may be important there. 
In this scenario, two nearby atoms experience an effective attraction since the 
substrate can deform in a cooperative manner, increasing the dimer's binding energy. 
At a formal level, the normal modes of the C atoms of graphene include a branch, the so 
called flexural phonons, whose dispersion relation at long wave-length 
for an unstrained sample is proportional to the square of the wave vector and this should 
change the phonon mediated interaction between adsorbed atoms. 
While the magnitude of this effect on graphene is unknown it may be sufficient to enhance 
the cohesion of otherwise marginally bound states, like the $^3$He liquid.
Free-standing graphene is not perfectly flat at a finite temperature but there is a small 
undulation \cite{ref47b,ref47c}.  
However the period of the dominant ripples is of order of 100 \AA~ and we do not expect 
that this can have a significant effect on the phase behavior of the adsorbed particles.

A second difference between graphene and graphite is relevant to the case of a free-standing 
graphene layer. In that geometry, gas will adsorb on both the top 
and bottom sides of the graphene. This means there is an additional attractive interlayer 
interaction present. The magnitude of the effect was estimated 
by Bruch , Kim, and Cole \cite{new1}. The resulting energy shift is small ($\sim$ 5\% in energy) 
because of the large separation ($> 6$ \AA) between the 
``top'' and ``bottom'' films; this result is expected in view of the small difference in 
Eq.(\ref{eqgrp2}) arising from the graphitic layers at similar, and larger, distances.

There exists one intriguing qualitative effect of the 2-sided adsorption problem. 
Consider atoms which form a commensurate $\sqrt{3}\times\sqrt{3}$ R30$^o$ phase on the top side. 
On the bottom side, then, there exist two alternative possible commensurate structures for the film. 
In one case, the adatoms occupy the same sublattice 
of sites as on the top. In the alternative structure, atoms occupy instead a complementary 
subset of sites. The site degeneracy is three in the former case 
and six in the latter, so these two possibilities involve phase transitions of different symmetry. 
Which of the two cases is appropriate for He on graphene 
is not known and this problem is currently under investigation in our group. 
In the first case, the $q=3$ Potts model is still appropriate. In the second case 
a simple extension of Alexander's argument might suggest $q=6$ but a more thorough study is needed. 
We note that the nature of the phase transition of 
the Potts model depends on $q$; for $q>4$ it is of first order \cite{ref48}.

One of the more interesting problems to explore is the effect of the adlayer on the electrical 
conductivity and other transport properties in the graphene sheet.
Experimental data concerning this property exist thus far only for Kr and He on nanotubes \cite{conduct}.
While there have been many studies of such behavior involving chemisorbed films, there have 
been few controlled studies of the pressure and temperature 
dependence for physisorbed films. The latter presents a distinct advantage, insofar as the 
adatoms provide weak, tunable perturbations of the 
graphene's electron gas \cite{new2}. This weakness arises from the large equilibrium 
distance and small charge transfer of the film, implying that the 
electron-He atom interaction is much smaller than occurs in chemisorption. 
Some preliminary experimental result on the electric conductivity of 
graphene as function of adsorption of Ar and He atoms has been reported \cite{ref49b}. 
This remains one of the interesting domains of future research in this field.

\subsection{Calculations for monolayer quantum films on graphene}

Since the potential energy on graphene differs from that on graphite only by a ($\sim$ 10-15\%) 
uniform energy shift, the 2D phase behavior is expected 
to be essentially the same on these surfaces. Indeed, that expectation has been confirmed thus 
far in the numerical studies of quantum films on graphene. 
However, there exist subtle, but qualitative, differences between these computations due to 
differences in the adsorption potential models and the simulated 
system size, as well as differences between studies using canonical and grand canonical 
simulation methods. For example, the use of an anisotropic He-C pair 
potential (motivated by He scattering experiments) gives rise to a significantly more 
corrugated potential than results from an isotropic pair potential. 
This can be quantified by
writing the adsorption potential \cite{ref49b0} $V(\vec{r})$ as a 2D Fourier expansion in the surface-parallel plane,
\begin{eqnarray}\label{ccpot}
V(\vec{r}) = V_0(z) + \sum_{\vec{G}\neq 0} V_{\vec{G}} (z) \exp^{i{\vec{G}}\cdot\vec{R}} \quad .
\end{eqnarray}
Here $\vec{R}=(x,y)$ and $\vec{G}$ is a 2D reciprocal lattice vector associated with the 
periodicity of the surface. An $\sim$ 1/3 larger amplitude of 
$V_{\vec{G}}(z)$, the corrugation potential, with anisotropy included (vs without anisotropy) 
was inferred \cite{ref49ba} from the matrix elements observed \cite{ref49bb}
in scattering between selectively adsorbed states of atoms which are
resonant with the incident beam. 
Such a large difference in the corrugation yields a significant difference in the 
energetics of the commensurate phase.

Gordillo, Cazorla and Boronat \cite{ref51,ref52,ref53,ref53b} explored the phase diagrams at 
$T=0$ of both quantum gases on graphene, finding little qualitative 
difference from the behavior known to occur on graphite. Their comparisons included that 
between the $\sqrt{3}\times\sqrt{3}$ R30$^o$ phase and a hypothetical liquid phase, 
conjectured by Greywall and Busch from heat capacity experiments. However, such a comparison 
suffers from the use of an isotropic He-C pair interaction, 
which favors the liquid phase, relative to the C phase. One of the findings of their study 
was the presence of a minute ($\sim$ 0.7\%) supersolid component in the case of 
commensurate $^4$He, which is absent from the corresponding state of para-H$_2$. 
They also report a significant superfluid fraction 
(up to 14\%) in $^4$He, at $T=0$, when vacancies are present; the vacancies' mobility is 
responsible for the large computed super-component. 
The finite value of the superfluid fraction for the perfect $\sqrt{3}\times\sqrt{3}$ R30$^o$ phase 
has to be regarded with caution because the 
quantum simulation method used (diffusion Monte Carlo) suffers from convergence 
properties \cite{ref49c} as function of the size of the so- called population 
of random walkers. In addition, the authors also found a finite superfluid fraction for 
the $\sqrt{3}\times\sqrt{3}$ R30$^o$ state of $^4$He on graphite and this finding is 
contradicted by experiment \cite{reppy}.

Kwon and Ceperley \cite{ref54} also explored the $^4$He phase on graphene at finite $T$ with 
PIMC, with a focus on the 
effect of varying the adsorption potential. Their results are consistent with those of Gordillo, 
Cazorla and Boronat, when the isotropic potential 
is used, but no superfluidity is found at the temperature of the simulation ($T=0.31$ K). 
A very different behavior was found when anisotropy is included. 
Specifically, they saw no evidence of a superfluid or a supersolid phase in the monolayer 
on graphene and vacancies in the $\sqrt{3}\times\sqrt{3}$ R30$^o$ 
state are localized. Their results include the analysis of the role of domains, as well as 
the presence of a high order commensurate phase, 
with fractional occupation 7/16 of the hexagonal sites. Similar findings are reported by 
Happacher et al. \cite{ref55} at temperature 0.5 and 1 K, 
even with an isotropic  He-C pair potential. The latter authors used a particularly large 
simulation sample and the grand canonical ensemble PIMC, 
rather than the canonical ensemble used by previous authors.

\section{Adsorption potentials on new materials derived from Graphene:
Graphane and Fluorographene}\label{sec4}

\begin{figure}[h]
\begin{center}
\includegraphics*[width=12cm]{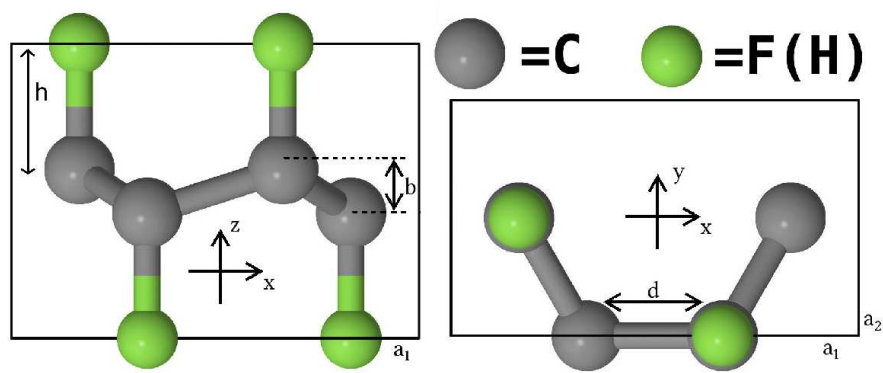}
\caption{(Color online) Geometry of the substrate with the definitions of the buckling parameter $b$, 
the interplane distance $h$ and $d$, the carbon--carbon distance projected
on the $x$--$y$ plane}\label{fg:fig1}
\end{center}
\end{figure}

\begin{figure}[h]
\begin{center}
\begin{minipage}{9cm}
\includegraphics*[width=9cm]{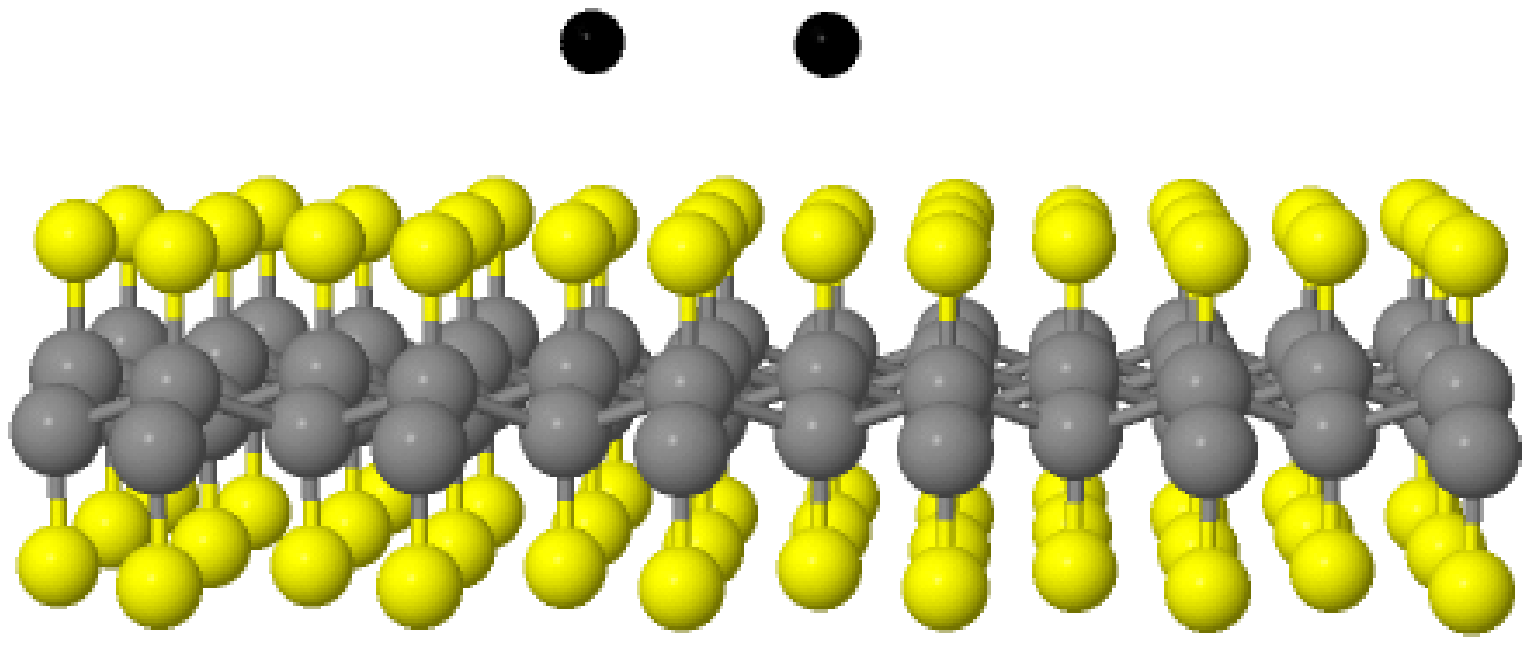}
\end{minipage}\hspace{1cm}
\begin{minipage}{4cm}
\includegraphics*[width=4cm]{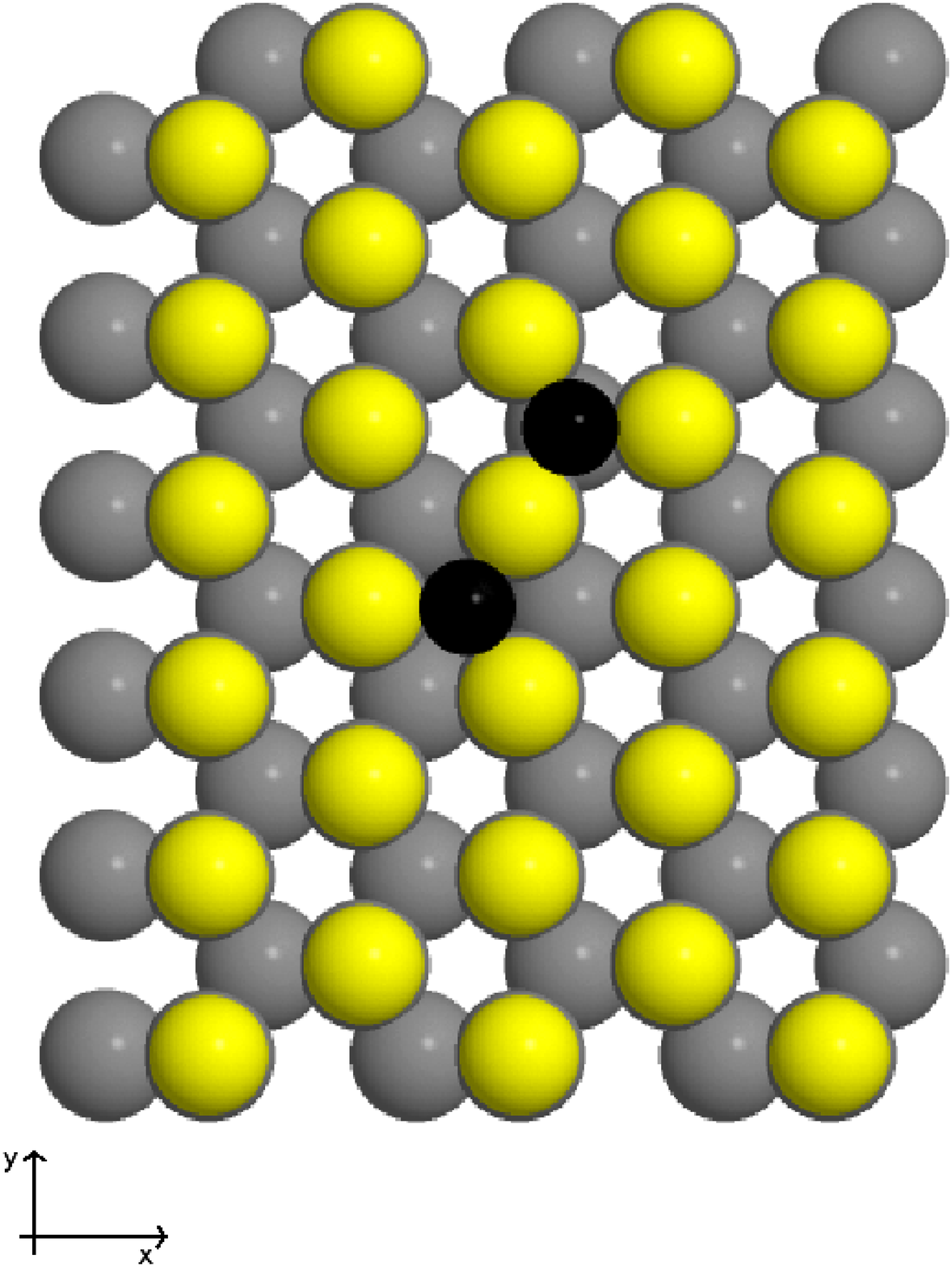}
\end{minipage}\caption{\label{fg:fig2} (Color online)
Two schematic views of GF. F (C) atoms are light (dark) gray. Positions of atoms
are to scale but their sizes are arbitrary. The black balls represent two
adsorption sites for He, one of each kind. GH is similar.
}
\end{center}
\end{figure}

Following the discovery of graphene \cite{refgraphene}, graphane and graphene--fluoride 
were considered as new chemical compounds on which respectively hydrogen and fluorine are bound on both 
sides of the graphene plane.
Both have been recently synthesized \cite{ref14,ref15}.
Graphane and graphene--fluoride have a similar geometry; half of the F (H) 
atoms are attached on one side of the graphene sheet to the carbon atoms forming 
one of the two sublattices of graphene. 
The other half are attached on the other side to the C atoms forming the other sublattice. 
The F (H) atoms are located on two planes (see left of Fig.~\ref{fg:fig2}); 
one is an overlayer located at a distance $h$ above the pristine graphene plane 
while the other is an underlayer at a distance $h$ below the graphene plane. 
In addition, as seen in Fig.~\ref{fg:fig1} there is a buckling of the C--plane with the C atoms of one 
sublattice moving upward by a distance $b/2=0.225$~\AA~ while the other sublattice moves 
downward by the same amount. A He (pH$_2$) atom approaching GF (GH) from above will 
interact primarily with the F (H) overlayer, but it will interact also with 
the C atoms and the F (H) atoms of the underlayer. 

We have adopted \cite{prb_fluorografene} a traditional, semi--empirical model to construct the potential 
energy $V(\vec{r})$ of a single He atom (pH$_2$ molecule) at position $\vec{r}$ near a 
surface \cite{lt26:ref5,lt26:ref5b,ref44}. 
The potential is written $V(\vec{r}) = V_{\rm rep}(\vec{r})  + V_{\rm att}(\vec{r})$, a 
sum of a Hartree--Fock 
repulsion derived from effective medium theory, and an attraction, $V_{\rm att}(\vec{r})$, 
which is a sum of damped He (pH$_2$) van der Waals (VDW) interactions and the 
polarization interaction with the surface electric field. The first term 
is $V_{\rm rep}(\vec{r})=\alpha_{EM} \rho(\vec{r})$, where $\rho(\vec{r})$ 
is the local density of the electrons of the substrate and
the value of $\alpha_{EM}$ is discussed in Reatto et al. \cite{lt26_nostro}
and reported in Tab.~\ref{ad:tabpar}.

The attraction is a sum of contributions; for GF,
\begin{equation}\label{vpotte}
V_{\rm att}(\vec{r}) = V_{\rm F+}(\vec{r}) + V_{\rm gr}(\vec{r}) + V_{\rm F-}(\vec{r})
- \alpha_{pol} {\rm E}^{2}(\vec{r})/2 \quad .
\end{equation}
The right--most term is the induced dipole energy, where $\alpha_{pol}$
is the static polarizability of the adsorbate and $\vec{E}(\vec{r})$ is the electric field 
due to the substrate.
In the case of $^4$He, this term gives a minor contribution to the adsorption potential and 
can be safely neglected. Due to the larger value of $\alpha_{pol}$ of H$_2$,
that electrostatic contribution is taken into account for H$_2$.
The three VDW terms for GF originate from the F overlayer, 
the graphene sheet and the F underlayer, respectively. These terms are 
described by the attractive part of a Lennard--Jones potential,
\begin{eqnarray}
V_{\rm F+}(\vec{r}) =-\sum_{j}\Gamma\left(\left|\vec{r}-\vec{r}_j^{\:F+}\right|\right)
\frac{C_{6F}}{\left|\vec{r}-\vec{r}_j^{\:F+}\right|^6} \label{vdw_damp}\\
V_{\rm F-}(\vec{r}) =-\sum_{j}\frac{C_{6F}}{\left|\vec{r}-\vec{r}_j^{\:F-}\right|^6} \label{vdw2}\\
V_{\rm gr}(\vec{r}) =-\sum_{j}\frac{C_{6C}}{\left|\vec{r}-\vec{r}_j^{\:gr}\right|^6} \label{vdw1}
\end{eqnarray}
where the sum is over the Carbon or Fluorine positions; $C_{6C}$ and $C_{6F}$ are  
the  VDW coefficient of He (pH$_2$) with respectively Carbon and Fluorine  \cite{refvdw1}.
The factor $\Gamma(r)$ is a damping function \cite{tangtoennies} to avoid the $r^{-6}$ divergence 
when the adatom comes close to the overlayer.
In the case of GH one has to replace F with H in the above formulae.

\begin{table}[h]
\caption{\label{ad:tabpar} Parameters for the adsorption potential for pH$_2$ and He}

\begin{center}
\begin{tabular}{ | c | c |  c |}
\hline
  Parameter & Value for H$_2$ & Value for He\\
\hline
$C_{6F}$ & 16.38 $eV$\AA$^6$ & 4.2 $eV$\AA$^6$ \\
$C_{6H}$ & - & 1.21 $eV$\AA$^6$\\
$C_{6C}$ & 9.96 $eV$\AA$^6$ & 3.45 $eV$\AA$^6$ \\
$\alpha_{EM}$ & 95.58 $eV$\AA$^{3}$ & 53.94 $eV$\AA$^{3}$ \\

\hline

\end{tabular}
\end{center}
\end{table}

\begin{figure}[h]
\begin{center}
\begin{minipage}{8cm}
\includegraphics*[width=8cm]{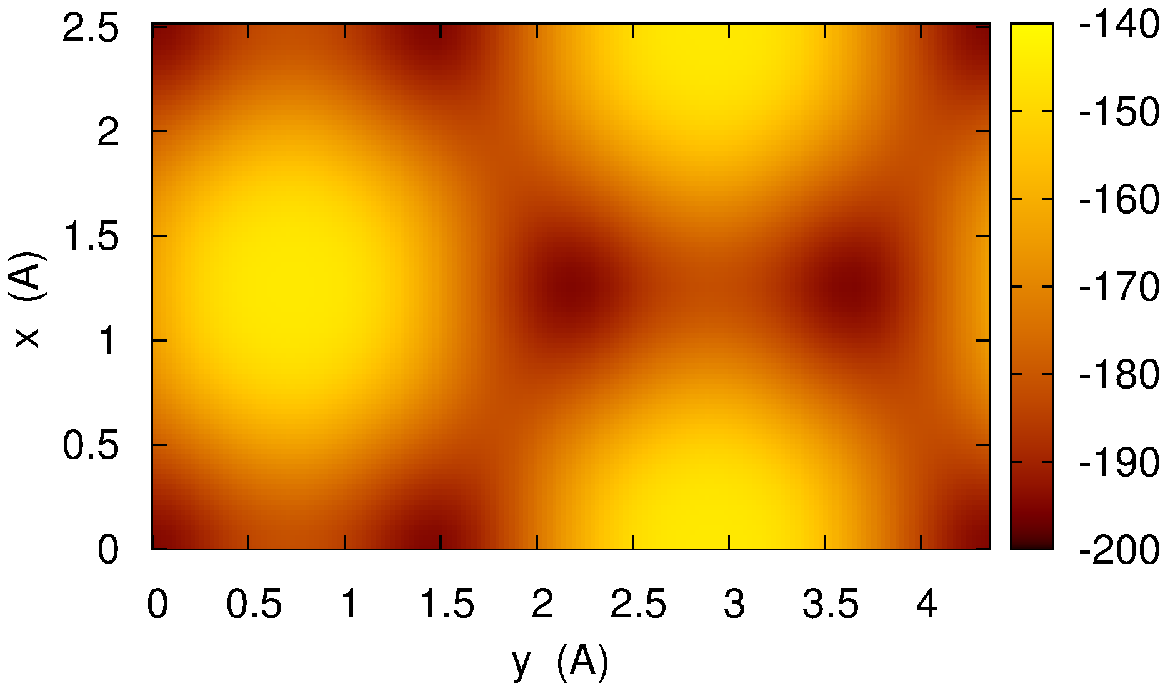}
\end{minipage}\hspace{1cm}
\begin{minipage}{6cm}
\includegraphics*[width=6cm]{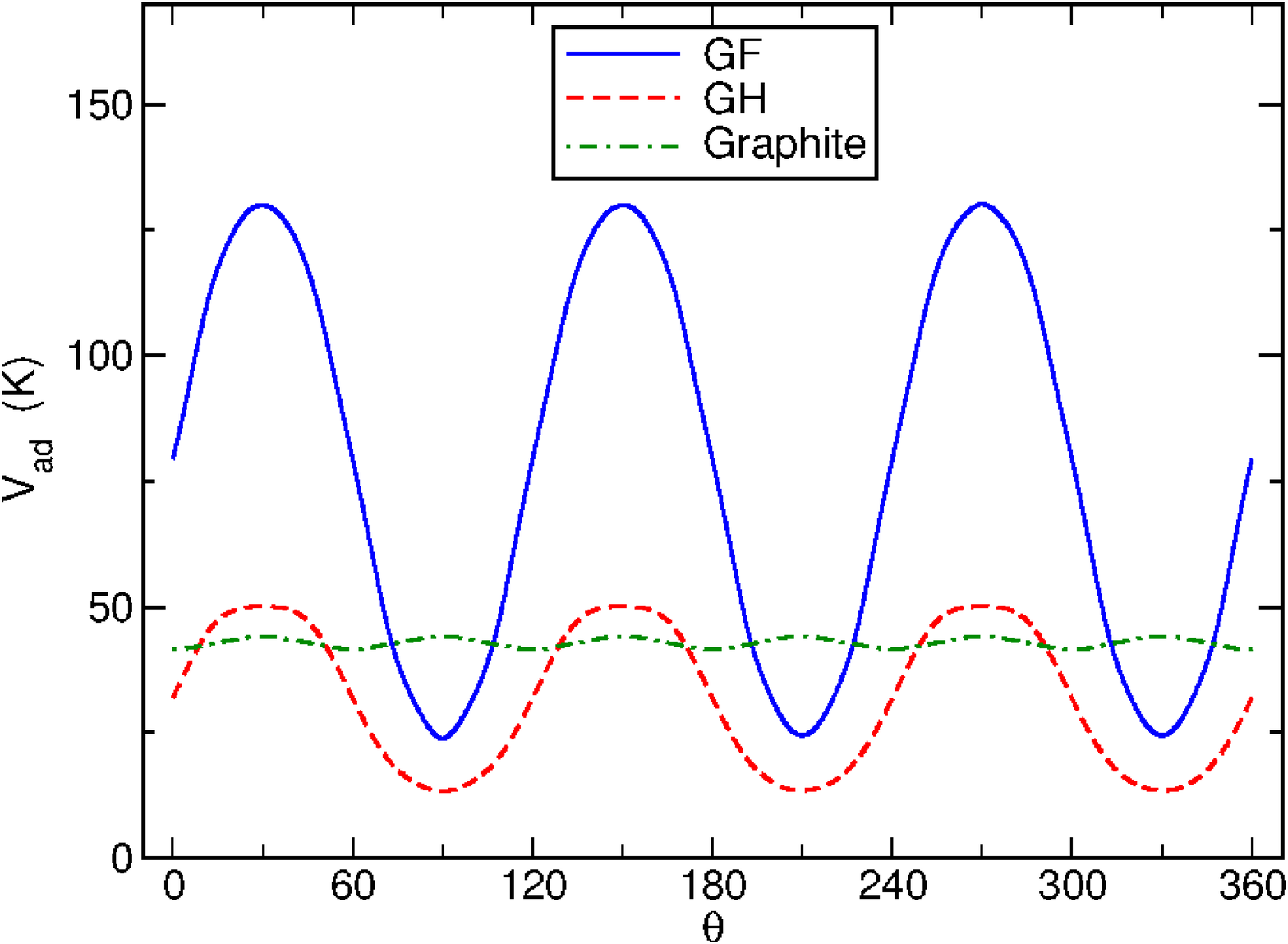}
\end{minipage}\caption{\label{ad:corrug} (Color online)
(Left) Plot of the minimum value with respect to $z$ of the adsorption potential He--GH in K as 
function of $x$--$y$.
(Right) Energy barrier in GF, GH and graphite as He atom moves along a line making an angle 
$\theta$ with the $x$ direction in the plane (see Fig.~\ref{fg:fig1})
and following the height $z(x,y)$ giving the minimum of $V\left({\vec r}\right)$.
Plotted energy is relative to energy at the adsorption site.
}
\end{center}
\end{figure}

For He atoms, the adsorption sites (see Fig.~\ref{ad:corrug}) are above the 
centers of each triplets of F (H) atoms of the overlayer, forming a honeycomb 
lattice with the number of sites equal to the number of C atoms, twice as 
many as those on Gr. Half of the sites are on top of hollow space (see Fig.~\ref{fg:fig2})
with no underlying atoms of the substrate and half of the sites
are above F (H) of the underlayer 
but the difference between the well depths for the two kinds of adsorption 
sites is very small, less than 1\%.
The well depth for GF is 498 K and for GH it is 195 K
(these values do not include the induced dipole energy which
gives a contribution below 1\% so it has been neglected)
whereas it is about 203 K for graphene.
The inter--site energy barrier is 24 K for 
GF and 13 K for GH. Both values are significantly smaller than the barrier 
height 41K for graphite. 
The attractive contribution of the F ion is large because of its charge.
While a similar contribution would be present on NaF, for example,
it would be absent from that on F2 crystal, for example. 

As shown in Fig.~\ref{ad:corrug}, the energy barrier on graphite does not
depend much on the direction in the $x-y$ plane whereas in the case of GF and
GH the ratio between maximum and minimum barrier height in the $x-y$ plane is
of order of 4--5: the energy landscape of the two last substrates is
characterized by a very large corrugation with narrow channels along which
low potential barriers are present. Another significant difference is that the
distance between two neighboring sites is $1.49$
\AA $\,$ for GF and $1.45$
\AA $\,$ for GH whereas it is $2.46$
\AA $\,$ for graphite and for Gr.
Prior to these studies, graphite was believed to be the most attractive
surface for He, with a well--depth a factor of 10 greater than that on the
least attractive surface (Cs) \cite{ref46}. The present results indicate that GF
replaces graphite, since its well is a factor of 3 more attractive.

The adsorption sites for H$_2$ on GF are the same as for $^4$He, 
the difference between the well depths of the two kinds of adsorption sites is of 45 K;
the lowest energy adsorption sites are in the hollow positions and the
well--depth is 2667 K. For comparison, on graphite the well-depth is 589 K.
The inter--site energy barrier on GF is 219 K for the lower minima and 174 K for the other kind; 
on graphite instead it is 37 K.

\section{Adsorption on Graphane and Fluorographene: predictions of novel phenomena for bosonic 
and fermionic atoms and molecules}\label{sec5}
The results presented in this section come from ``exact'' Quantum Monte Carlo simulations. 
In particular, 
the results at zero temperature have been obtained with the Path Integral Ground State 
method \cite{pigs} and those at 
finite temperature with the Path Integral Monte Carlo method \cite{ceperley}. 
Both methods express a quantum expectation value of a bosonic system like a multi--dimensional 
integral that can 
be computed exactly with the Metropolis \cite{metropolis} algorithm. 
A positive attribute of those methods 
is that they are ``exact'', in principle, where ``exact'' means that the error of the 
employed approximations can be arbitrarily reduced below the statistical uncertainty of 
the Monte Carlo methods.
Details on the computational methods can be found in Ref.~\onlinecite{prb_fluorografene}.

We have described in the previous section the properties of the adsorption potential 
between He (H$_2$) and GF (GH). Compared with the adsorption potential of He (H$_2$) on Graphite,
remarkable differences arise even at a qualitative level; 
the symmetry of the adsorption sites is different and the intersite distance is much smaller.
In the case of GF the substrate is more attractive than graphite.
In addition, the corrugation is stronger and
generates channels in which a helium
atom is more likely to cross from one adsorption site to another.
\begin{figure}[h] 
\begin{center}
\includegraphics*[width=16cm]{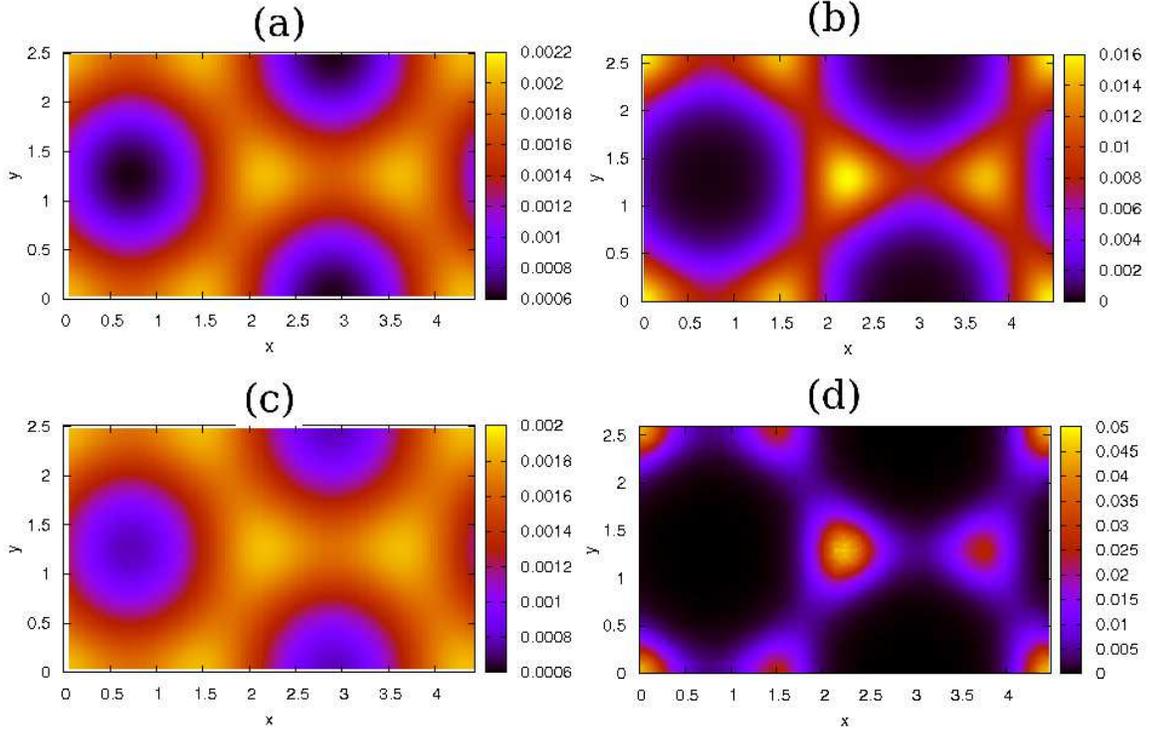}
\caption{\label{n1psi} (Color online)
Square modulus of the wave function (in \AA$^{-2}$ units) of a single atom (molecule), 
in the $x$--$y$ plane, integrated along the $z$ direction. 
Both GF and GH have been considered, and the plot  is shown for different atoms: 
(a) $^4$He on GH, (b) $^4$He on GF,  (c) $^3$He on GH, (d) H$_2$ on GF.
The axis labels are in \AA~ units.
} 
\end{center}
\end{figure}

\subsection{A single atom on GF and GH}
The previously mentioned features of the adsorption potential suggest that a helium 
atom (or H$_2$ molecule) on GF (GH) is able to 
tunnel quickly to neighboring adsorption sites and is thus delocalized over the substrate. 
This is indeed the case, as is shown in Fig. \ref{n1psi}, 
where the square modulus of the wave function of a single Helium atom (H$_2$ molecule) is 
represented. 

The boxes on the left in Fig.~\ref{n1psi} compares the two isotopes of Helium on GH; as expected 
from its larger zero point motion, $^3$He (c) has a square modulus wave function that is more 
delocalized than its bosonic counterpart $^4$He (a). The boxes on the right represents $^4$He (b) and 
H$_2$ (d) on GF; the main difference here is that the H$_2$-GF interaction is about five times 
more attractive than the He-GF one. As a consequence, for H$_2$ the channels that connect two 
adjacent adsorption sites become narrower and the localization around an adsorption minimum is larger. 
Another consequence of the larger energy scale 
of the H$_2$-GF interaction is that the two kinds of adsorption minima are no longer degenerate. 

\begin{figure}[h] 
\begin{center}
\includegraphics*[width=16cm]{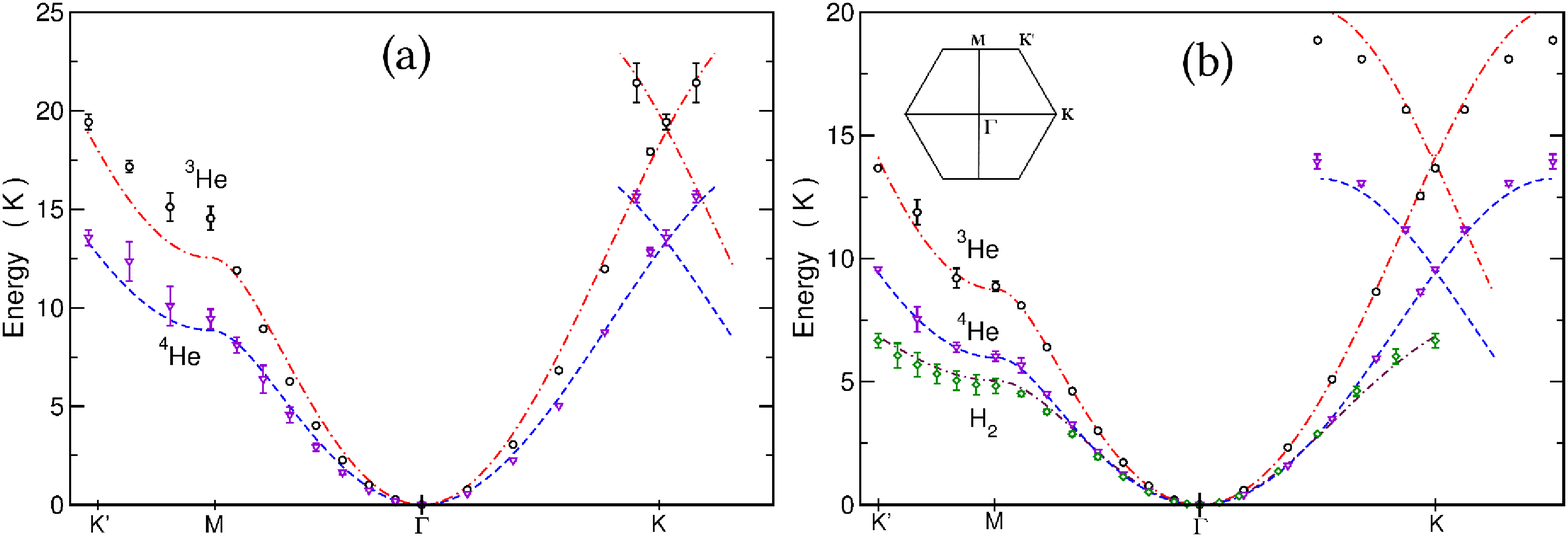}
\caption{\label{n1band} (Color online)
(a) First energy band of the two isotopes of Helium on GH, with a part of the second energy band 
around the Dirac point K. (b) 
The same of (a) on GF for the two isotopes of Helium and H$_2$. In the inset a schematic 
representation of the 
first Brillouin zone given by the periodicity of the substrate potential.
The dashed lines are fit to the data with tight binding model of 
a honeycomb lattice with nearest and next nearest neighbors coupling $t_1$ and $t_2$: 
$\epsilon({\bf k})=\pm t_1\sqrt{3+f({\bf k})}-t_2f({\bf k})$, where 
the plus(minus) sign is for the upper(lower) band and 
$f({\bf k})=2\cos(\sqrt{3}k_y a) + 4\cos(0.5\sqrt{3}k_y a)\cos(1.5k_x a)$ with $a$ the nearest 
neighbor distance.
In principle one should allow for different values of $t_2$ for the two adsorption sites
in the unit cell. This does not seem necessary given the goodness of the fit with a single value for $t_2$.
The error bars in this figure represent the uncertainty associated with the inversion of the 
Laplace transform required to obtain 
the excitation energies from the imaginary time dynamics.
} 
\end{center}
\end{figure}

In Fig. \ref{n1band} we show the lowest energy bands of both He isotopes on GF and GH and 
H$_2$ on GF; these results 
have been obtained from the inversion of the Laplace 
transform \cite{gift} of the density-density correlation function in imaginary time, a 
quantity computed exactly with 
Path Integral methods. 
The state of wave vector {\bf k} of the first energy band, in fact, has a dominant role 
in the density--density 
correlation function in imaginary time $\tau$, 
$F({\bf k},\tau)=\langle\rho_{\bf k}(\tau)\rho_{\bf k}(0)\rangle$, where 
$\rho_{\bf k}(\tau)=\exp^{i{\bf k}\cdot\hat{\bf r}(\tau)}$ and $\hat{\bf r}(\tau)$ is the 
position of the atom at imaginary time $\tau$.
From the first energy band the effective to bare mass ratio $m^{\star}/m$ can be computed. 
This quantity reflects the different corrugation of the 
adsorption potential and is shown in Table \ref{tabeffec}.

\begin{table}[h]
\caption{\label{tabeffec} (Left) Effective--to--bare mass ratio $m^{\star}/m$ of He and H$_2$ on 
different substrates, derived from the small wave-vectors behavior of the first energy band.
(Right) Nearest neighbor and next--nearest neighbor fitting parameters  $t_1$ and $t_2$ in 
Kelvin units for the tight binding model used in Fig.~\ref{n1band} for GF(GH).}

\begin{center}
\begin{minipage}{6cm}
\begin{tabular}{ | c | c |  c | c |}
\hline
  Adsorbate & GF & GH & GR\\
\hline
H$_2$  &  2.73 & - & 1.03 \\
$^3$He  &  1.25 & 1.01 & 1.08 \\
$^4$He  &  1.40 & 1.05 & 1.10 \\
\hline

\end{tabular}
\end{minipage}\hspace{1cm}
\begin{minipage}{6cm}
\begin{tabular}{ | c | c |  c |}
\hline
  Adsorbate & $t_1^{GF(GH)}$ & $-t_2^{GF(GH)}$ \\
\hline
H$_2$  &  1.95  & -0.24  \\
$^3$He  &  5.695(6.434) & 0.33(0.04) \\
$^4$He  &  3.63(4.514) & 0.16(0.021)  \\
\hline

\end{tabular}

\end{minipage}
\end{center}
\end{table}

\begin{figure}[h] 
\begin{center}
\includegraphics*[width=10cm]{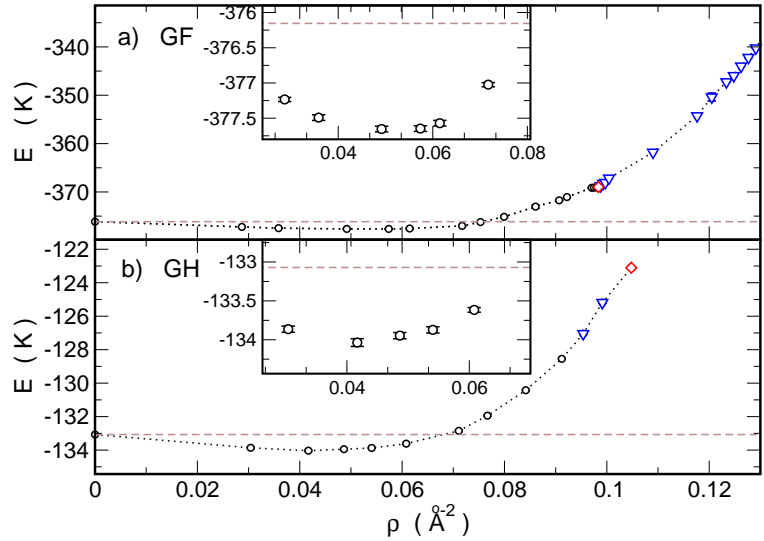}
\caption{\label{neqeos} (Color online)
Energy per atom of $^4$He on GF (a) and GH (b) at $T=0$ K with an inset showing in detail the 
density region around the energy minimum. 
Circles represent liquid phases, down triangles are solid incommensurate phases and the square 
indicates the commensurate phase at filling factor $x=2/7$. 
The horizontal dashed line is at the level of the energy of a single atom on the substrate.
Error bars (not shown) are smaller than the symbols.
} 
\end{center}
\end{figure}
\subsection{Monolayer film of $^4$He and $^3$He on GF and GH}
\begin{figure}[h] 
\begin{center}
\includegraphics*[width=10cm]{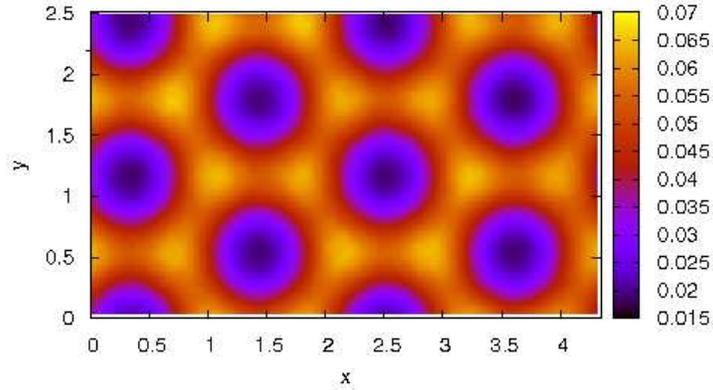}
\caption{\label{neqdens} (Color online)
Local density in \AA$^{-2}$ units for a system of $N=41$ atoms of $^4$He on GH at equilibrium 
density and 
at $T=0$ K. The axes are labelled in \AA~units.
} 
\end{center}
\end{figure}

When a monolayer of $^4$He atoms interacting with an HFDHE2 potential \cite{aziz79} and adsorbed 
on GF (GH) is considered,
the energy per particle $E(\rho)$ has a smooth dependence on density $\rho$ with a minimum at
a finite density, the equilibrium density.
Therefore the ground state of $^4$He on GF and GH is a 
{\it self--bound modulated superfluid} of density $\rho_{eq}^{GF} = 0.049 $~\AA$^{-2}$ for GF and 
$\rho_{eq}^{GH} = 0.042 $~\AA$^{-2}$ for GH. For comparison, the strictly 2d system \cite{whitlock}
 of $^4$He 
has an equilibrium density of $\rho_{eq}^{2d} = 0.0436$~\AA$^{-2}$ which is surprisingly similar.
This is different from the case of the monolayer of $^4$He on Graphite, where the 
equilibrium density is the $\sqrt{3} \times \sqrt{3}$ R30$^o$ solid commensurate phase. 
A commensurate state similar to that on graphite could be present also on GF(GH) at density
$\rho = 0.0573$~\AA$^{-2}$ ($\rho = 0.0608$~\AA$^{-2}$) at filling factor 1/6 of the
adsorption sites (the different values of density are due to different lattice expansions
relative to graphite).
A scaling analysis \cite{lt26_nostro} 
of the static structure factor $S({\bf k})$ at this coverage as function of the size of the system
showed that there is no signature of such a commensurate phase for $^4$He on GF (GH),
i.e. the corresponding Bragg peaks are absent
in the static structure factor; this ordered phase is thus unstable on 
GF (GH). As shown in the equation of state at $T=0$ K in Fig.~\ref{neqeos}, the equilibrium ground 
state density is below that of the hypothetical commensurate phase similar to
the $\sqrt{3} \times \sqrt{3}$ R30$^o$ of graphite. 
Fig.~\ref{neqdens} shows the local density on the $x$--$y$ plane for $^4$He on GH. 
\begin{figure}[h]
\begin{center}
\includegraphics*[width=8cm]{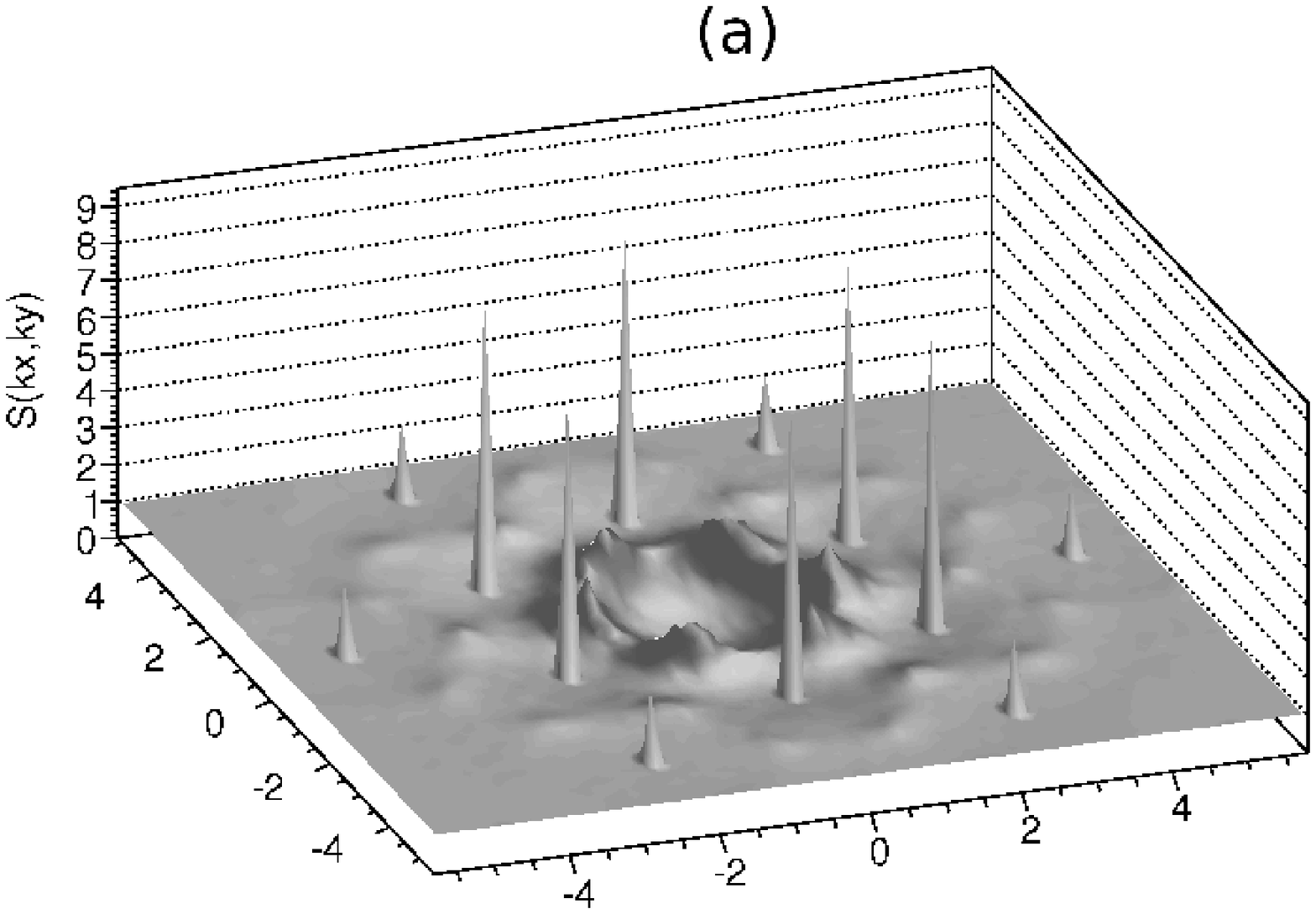}
\includegraphics*[width=7cm]{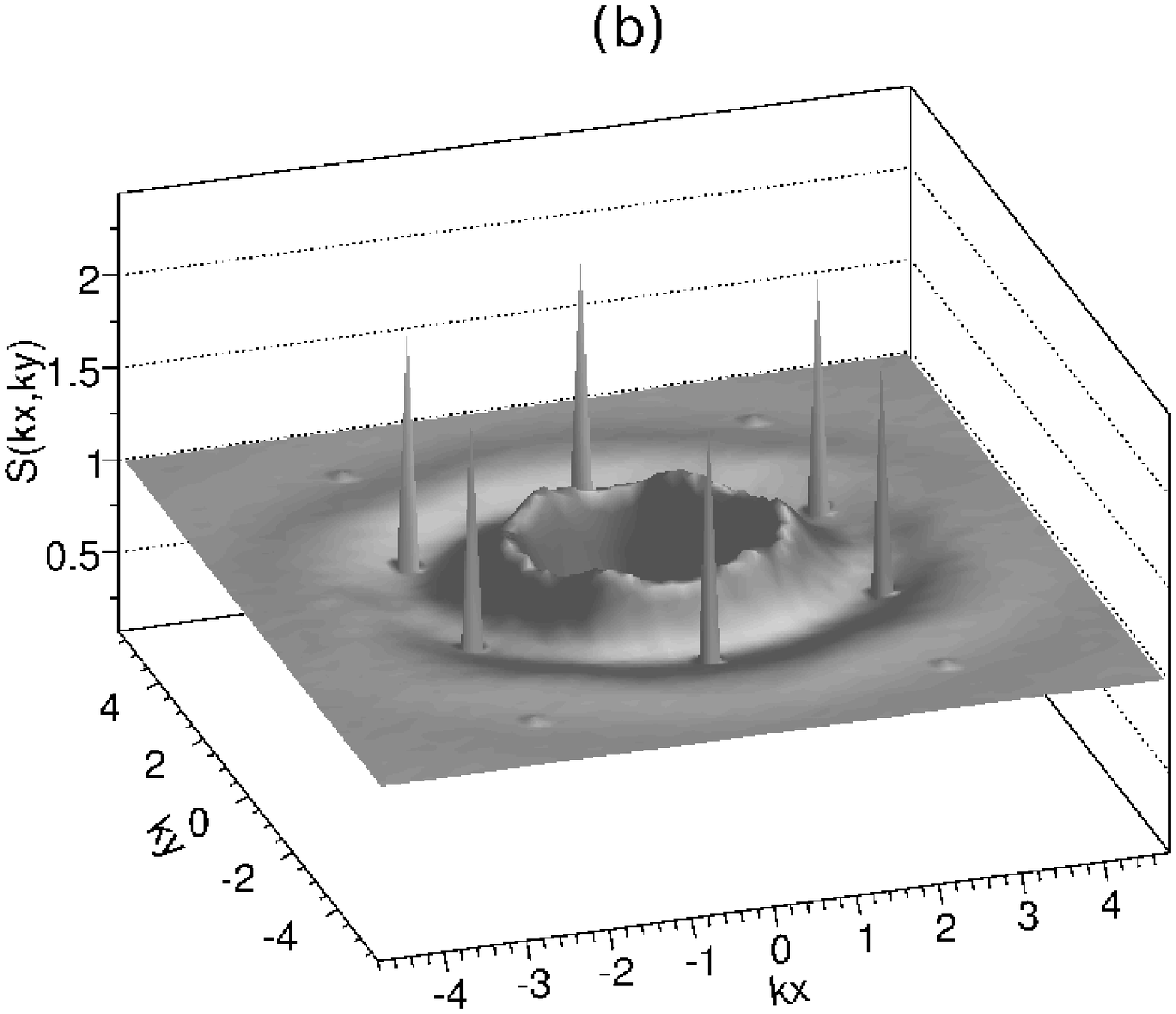}
\caption{\label{neqsdk}
Static structure factor on the $k_x$--$k_y$ plane in the reciprocal lattice of $N=96$ atoms of $^4$He 
on GF (a) and on GH (b) at equilibrium density.
The six higher peaks represent the density modulation induced by the substrate potential.
} 
\end{center}
\end{figure}
In this figure, clearly visible is the highly modulated local density imposed on the monolayer by the 
substrate potential. The modulation in density is reflected in Fig.~\ref{neqsdk}, where 
the static structure factor on the $k_x$--$k_y$ plane of the reciprocal space has been shown for 
both GF and GH: 
we note here the {\it crater--like} signal at small wave--vectors that is characteristic of short
range order of a fluid phase as well 
as the presence of sharp peaks at wave vectors that correspond to the expected density modulation 
induced by the substrate. 
The static structure factors are consistent with the presence of a liquid phase at equilibrium 
density, this 
evidence is confirmed by a scaling analysis of the monolayer that has been presented in 
Ref.~\onlinecite{prb_fluorografene}, 
along with the computation of the one--body density matrix and superfluid fraction. 
The $T=0$ K condensate fraction $n_0$ at 
equilibrium density is 11.1(1)\% for GF and  22.6(13)\% for GH. These values are much smaller 
than the strictly two dimensional 
case \cite{whitlock}, for which $n_0 \sim 0.4$; this is a consequence of the spatial modulation 
induced by the substrate as well as the 
smaller effective surface available to the atoms. The $T=0$ K superfluid fraction $\rho_s/\rho$ is 0.60(3) 
for GF and 0.95(3) for GH.
The fact that the superfluid density is not 1 at $T=0$ K was indeed predicted \cite{new0a,new0b}
for such kind of adsorption systems as an effect of the corrugation of the external potential.
The superfluid fraction has been computed also at finite $T$ and we have estimated \cite{prb_fluorografene}
that the superfluid--normal fluid transition is 0.2-0.3 K for GF and 1.0-1.2 K for GH.

\begin{figure}[h]
\begin{center}
\includegraphics*[width=12cm]{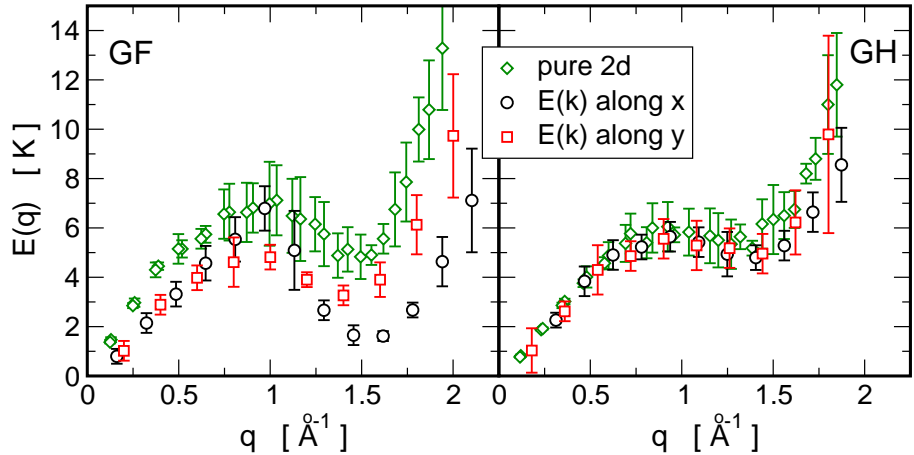}
\caption{\label{neqspec} (Color online)
Low energy excitation spectrum of $^4$He on GF and on GH at equilibrium density and $T=0$ K.
For $^4$He on GF, note the different energies of
the maxon and the roton states along the $x$ and $y$ directions; this anisotropy becomes
smaller than the error bar in the
phonon region located at small wave--vectors. The open diamonds represent the
excitation spectrum of pure 2d $^4$He at a density $\rho=0.049$~\AA$^{-2}$ in the GF case
and $\rho=0.0421$~\AA$^{-2}$ in the GH case.
}
\end{center}
\end{figure}
We have also computed \cite{jltpnostro} the excitation spectrum of the system from the density 
correlation function for imaginary
time and using a powerful inversion method to real frequencies \cite{gift}.
In Fig.~\ref{neqspec} the low--energy excitation spectrum 
at the equilibrium density is shown for $^4$He on both GF and GH. The spectrum is of 
{\it phonon--maxon--roton} type, 
typical of bulk superfluid $^4$He,
but the energy of the roton is anisotropic with respect of the direction 
of the wave--vector;
this novel feature is dramatic in the case of GF. For GH, in contrast, the anisotropy is smaller 
than the precision of the 
computation. 
It is interesting to compare these excitation spectra with that of $^4$He in mathematical $2d$
at the same density.
The spectrum in $2d$ has been computed recently \cite{arrigoni},
and it is shown in Fig.~\ref{neqspec} at the densities of the adsorbed film.
It can be seen that the spectrum of $^4$He on GH is very similar to that in $2d$.
This probably reflects the fact that the $^4$He local density on GH is modulated by the substrate 
but it has a sizeable
value even at the unfavorable positions on top of the H atoms as can be seen in Fig.~\ref{neqdens}.
The situation is quite different for $^4$He on GF:
obviously the $2d$ model does not capture the anisotropy of the rotons on GF but also the roton 
energy of the $2d$ model is much larger (left panel of Fig.~\ref{neqspec}).
This can be understood by the fact that the local density of $^4$He
on top of the F atoms is essentially zero so that it is as though the $^4$He atoms are moving in a 
multiconnected space.

\begin{figure}[h] 
\begin{center}
\includegraphics*[width=15cm]{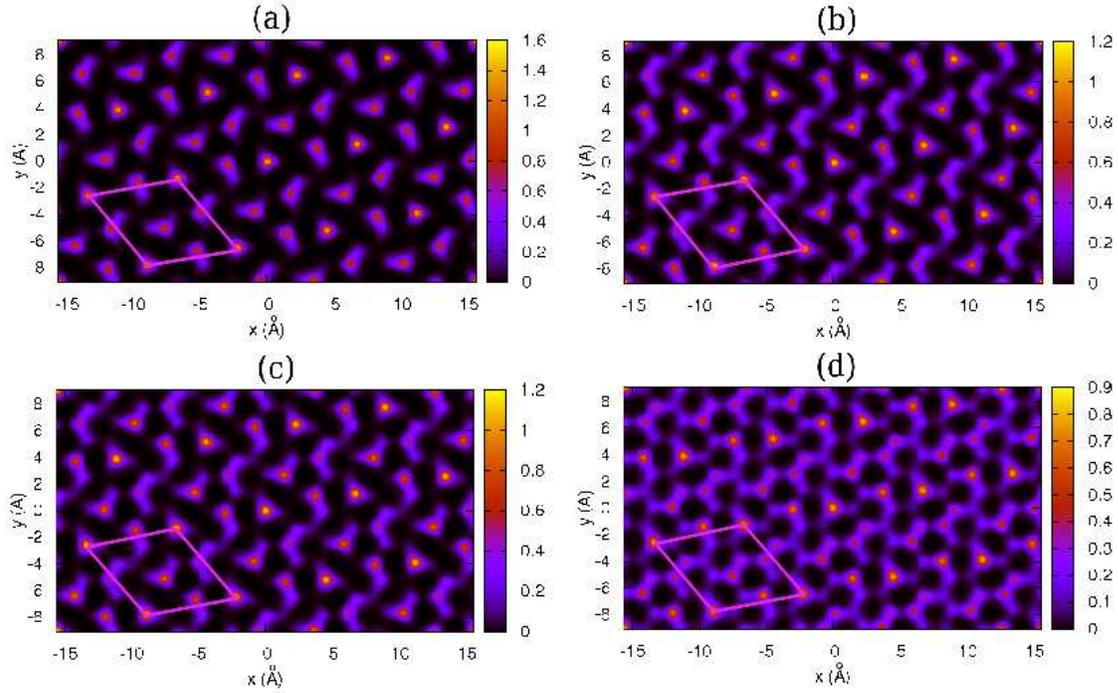}
\caption{\label{nsoldens} (Color online)
$^4$He local density, $\rho(x,y)$, in ~\AA$^{-2}$
(color code to the right of each panel) at four different temperatures
on the $x$--$y$ plane of the $2/7$ commensurate
phase of $^4$He on GF.
The temperatures are (a) 0 K, (b) 0.75 K, (c) 1 K and (d) 1.5 K. The unit cell is marked in 
all the figures with 
a dark pink line.
} 
\end{center}
\end{figure}

At higher densities of the first monolayer, on both GF and GH a triangular commensurate $2/7$ phase
is found to be stable. The $2/7$ phase consists of
a triangular lattice rotated by 19.1$^o$ with respect to the substrate potential; this lattice 
is obtained 
by placing 2 atoms every 7 adsorption sites. In the unit cell (see Fig.~\ref{nsoldens}) one 
of the $^4$He atoms is localized on an adsorption site 
above a triplet of F(H) atoms,
other two atoms approach adsorption sites of the other kind and finally the
fourth one is centered on a saddle point of the potential. 

\begin{figure}[h] 
\begin{center}
\includegraphics*[width=10cm]{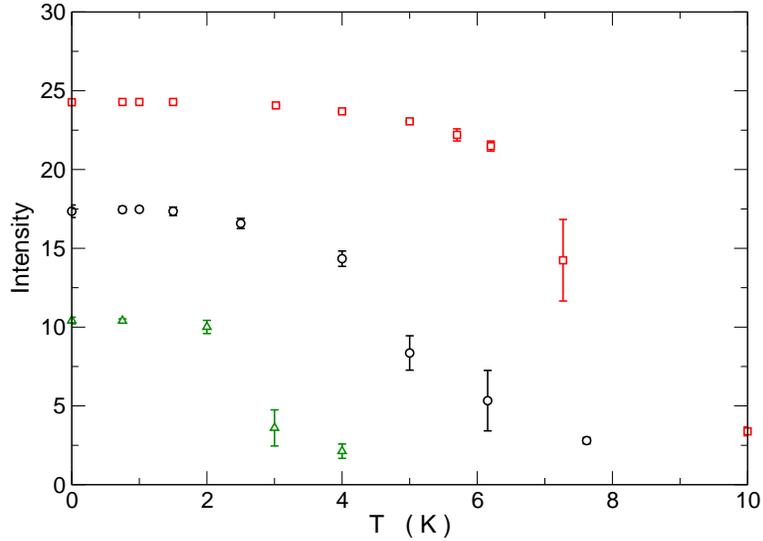}
\caption{\label{nsolbragg}
Temperature dependence of the average intensity of the main Bragg peaks characteristic of the $2/7$ 
commensurate phase of $^4$He 
on GF (Circles) and of $^4$He on GH (Squares). (Triangles) The same for the $1/3$ commensurate 
phase of $^4$He on Graphite.
} 
\end{center}
\end{figure}

This commensurate phase is analogous to the 
$4/7$ phase found on the second layer of $^4$He on Graphite \cite{ref_47he3}.
The local density of the $2/7$ phase for $^4$He on GF is shown in Fig.~\ref{nsoldens} for different 
temperatures $T$; in this phase, not all atoms are localized around a single adsorption site:  
some atom visits two or three neighboring sites; as consequence, the atoms are rather mobile 
and exchange easily, but there is still spatial order; this particularity might give rise to 
new phenomena and indeed this system is a good candidate for the possibility of supersolidity.
Figure \ref{nsolbragg} shows the average intensity of the 
Bragg peaks characteristic of the $2/7$ as a function of the temperature $T$. 
For purpose of comparison the result for $^4$He an graphite at 1/3 coverage is also shown.

The effect of the temperature introduces particularly large disorder in this commensurate phase
of the monolayer.
In this context the order is measured by the intensity of the Bragg peaks of 
the $2/7$ phase, which decreases with increasing temperature until it reaches an intensity 
comparable with that typical of the short range of the liquid phase.
The $2/7$ phase is ordered up to a temperature of roughly 6 K for GF and 7 - 8 K for GH; however, 
for GH the promotion of some atom to the second layer occurs at a lower temperature, roughly at 6 K, 
whereas for GF this promotion occurs between 10 - 20 K.

At coverages around 2/7 we found that $^4$He has an incommensurate triangular order deformed by the 
substrate potential and defective because
such order is not compatible with the periodic boundary contitions imposed at the edges of the 
simulation box. 
The static structure factor typical of an incommensurate phase of $^4$He on GF (GH) (not shown 
here, see Ref.~\cite{prb_fluorografene}), 
has a contribution that comes from a star of six peaks as expected for a triangular solid. 
The wave vectors of these peaks are close to those expected for 
an ideal triangular solid at the same density; also the modulus of the peaks increases in a 
smooth way as the density is 
increased, following the behavior of a triangular solid. The observed deviation of the wave 
vectors from the 
value $k_{Bragg} = 4\pi\left(\rho/2\sqrt{3}\right)^{1/2}$ of an ideal 
triangular solid arises from the deformation of the lattice and also from the 
presence of some defects, mainly dislocations, that can be observed from some sampled 
configurations of the atoms.

Increasing the number of particles in the monolayer, at some point, some atoms begin to be 
promoted to the second layer.
At $T=0$ K this first takes place at a density 
$\rho_{sat}^{GF}=0.136$~\AA$^{-2}$ for the GF case and a density 
$\rho_{sat}^{GH}=0.108$~\AA$^{-2}$ for the GH case; the higher value for GF reflects the 
higher binding energy on this surface. 
Beyond such densities, the occupation of the second layer manifests itself as an increasing 
secondary peak in 
the local density along the $z$--direction.

Monolayers of the fermionic $^3$He at low coverage on GF and GH have been studied \cite{prb_fluorografene}
with a strategy similar to that adopted in Ref.~\onlinecite{ref38c} for $^3$He in 2d.
One starts from Bosons of mass 3 and computes the energy gap between the ground state
of the Fermion system and that of the Boson system from a suitable correlation function
in imaginary time.
For both substrates no commensurate state is found to be stable at low coverage but $^3$He
atoms are found in an anisotropic fluid phase. In the case of GH, no self bound liquid phase
is found, i.e. at all finite densities the energy per particle is above the adsorption
energy of a single atom.
In the case of GF the energy vs $\rho$ has a minimum at a finite density so that the ground
state should be a self--bound liquid.
However, the minimum is very shallow such that it is within the precision of the computation,
so further computations are needed in order to confirm these results.
What is firmly established is that $^3$He on GF and GH gives the possibility of studying
a new {\em anisotropic} Fermi fluid with tunable density.

The monolayer films of $^4$He and $^3$He on GF and GH have thus been proved to exhibit novel physical 
properties. They are indeed  
complex systems with rich phase diagrams that are currently under study. In particular, 
the predicted commensurate phases of $^4$He 
are remarkable and very promising candidates for research concerning both superfluidity 
and Bose-Einstein condensation in solid systems.
We have already remarked that our adsorption potential of He on GF and GH is affected by
a degree of uncertainty. It is important to notice that the reported results are roboust to details
of the adsorption potential. In fact we have verified \cite{prb_fluorografene} that the qualitative results do not
change when the less certain interaction parameters in Eq.~\ref{vpotte} have been modified by up to $\pm 20$\%.

\section{Summary: the future of this field}\label{sec6}

This review has described a variety of problems involving quantum films on surfaces derived from 
graphite or graphene. 
In the case of a graphite surface, most of the monolayer phase behavior is understood 
semiquantitatively. By the word ``understood'', 
we mean that the phases observed experimentally are, overall, in relatively good agreement 
with theory or computer simulations. 
This satisfactory situation represents a significant accomplishment because of the 
experimental and theoretical complications presented by quantum films weakly 
interacting with the substrate (compared to heavier, more strongly interacting gases). 
For example, many traditional surface science probes 
(STM, LEED, inelastic scattering) are not appropriate for these films because they are 
either too strongly perturbing or because the film is too weakly interacting. 
On the theoretical side, the challenge of strong interadsorbate interactions and quantum fluctuations 
render the analysis particularly challenging, especially for the Fermi system $^3$He.

The cases of quantum gas adsorption on graphene, fluorographene and graphane have received 
intense attention recently, from theory and simulations, 
but relatively little experimental effort. This situation is unfortunate because some 
remarkable properties have already been predicted for these systems, 
especially in the $^4$He case. Most dramatic, perhaps, is the prediction of a highly 
anisotropic superfluid monolayer with superfluid fraction 
$\rho_s/\rho \sim 0.6$ at $T=0$ and anisotropic roton excitations. 
Such a film state has never been observed in the laboratory in more than a half-century of study. 
Another point of interest is the prediction of a commensurate state at 2/7 coverage on GF and GH. 
In this state only 1/4 of the adsorbed $^4$He atoms are well 
localized at adsorption sites, the other atoms do not occupy specific adsorption sites but 
tunnel among different sites. This appears a possible candidate for a supersolid state.

After completion of this topical review two studies have been published by Boronat, Gordillo and collegues
\cite{giapin,giapin2}.
In the first they have investigated the role of anisotropic pair potential for He on both graphene 
and a carbon nanotube. In the second they have studied the behavior of H$_2$ on GH at $T=0$. 
The main finding is that the ground state is a commensurate solid with order similar to the 
$\sqrt{3}\times\sqrt{3}$ one, found for H$_2$ and D$_2$ on graphene \cite{ref51,ref53b}.
Subsequent to this paper's submission we received a preprint (now published) of a study by
Vranje\v{s} Marki\'c and collegues \cite{preprint} concerning helium droplets adsorbed on graphene.
That specific topic is not within the domain of the present review.
We mention the paper, however, because it provides a detailed discussion of the numerous 
uncertainties in the adsorption potential, a limitation relevant to all calculations involving adsorption.
Only extensive experimental study can help to resolve these uncertainties.

While this review has focused on the monolayer film properties per se, there are numerous 
fundamental aspects of the broader problem which should stimulate 
our community to explore this class of systems. One example involves the fact that a 
physisorbed film represents a controllable, comprehensible, 
weak perturbation of the electronic properties of the substrate. 
For graphene (as well as for nanotubes), the latter properties are remarkable, 
resulting already in one Nobel Prize, so probing with quantum films is potentially invaluable.

Summarizing, this field is blossoming in many directions and we look forward to a coming decade 
of exciting scientific discovery. 

\section{Acknowledgements}
We would like to acknowledge helpful correspondence with Hiroshi Fukuyama,
who kindly provided a figure, and discussions with Tony Novaco and Oscar
Vilches.

This work has been supported by Regione Lombardia and CINECA Consortium
through a LISA
Initiative (Laboratory for Interdisciplinary Advanced Simulation) 2012
grant [http://www.hpc.cineca.it/services/lisa].

\end{document}